\documentclass[12pt]{iopart}
\usepackage{iopams}

\usepackage[comma, authoryear]{natbib}
\usepackage{float}
\usepackage{my_amsmath}
\usepackage{amssymb}
\usepackage{bm} 
\usepackage{textcomp}

\usepackage{graphicx}
\usepackage{caption}
\usepackage{float}

\usepackage{array}
\newcolumntype{P}[1]{>{\centering\arraybackslash}p{#1}}
\newcolumntype{M}[1]{>{\centering\arraybackslash}m{#1}}
\usepackage{tabularx}

\def\rd{\boldsymbol{\mathfrak{r}}}
\mathchardef\mhyphen="2D

\usepackage{hyperref}

\usepackage{makecell}

\usepackage[author={Ziping Liu}]{pdfcomment}
\usepackage[rgb]{xcolor}

\begin{document}

\begin{titlepage}

This manuscript has been accepted for publication in Physics in Medicine and Biology. 
Please use the following reference when citing the manuscript.

Liu, Z., Mhlanga, J.C., Laforest, R., Derenoncourt, P.R., Siegel, B.A. and Jha, A.K., 2021. A Bayesian approach to tissue-fraction estimation for oncological PET segmentation. Physics in Medicine \& Biology, \textbf{66}(12), p.124002.

\end{titlepage}

\clearpage

\maketitle
\title{A Bayesian approach to tissue-fraction estimation for oncological PET segmentation}

\author{Ziping Liu\textsuperscript{1}, Joyce C. Mhlanga\textsuperscript{2}, Richard Laforest\textsuperscript{2}, Paul-Robert Derenoncourt\textsuperscript{2}, Barry A. Siegel\textsuperscript{2} \& Abhinav~K.~Jha\textsuperscript{1,2}}

\address{$^1$Department of Biomedical Engineering, Washington University in St. Louis, St. Louis, MO, 63130, USA}
\address{$^2$Mallinckrodt Institute of Radiology, Washington University School of Medicine, St. Louis, MO, 63110, USA}
\ead{a.jha@wustl.edu}

\begin{abstract}
Tumor segmentation in oncological PET is challenging, a major reason being the partial-volume effects that arise due to low system resolution and finite voxel size. The latter results in tissue-fraction effects, i.e.~voxels contain a mixture of tissue classes. Conventional segmentation methods are typically designed to assign each voxel in the image as belonging to a certain tissue class. Thus, these methods are inherently limited in modeling tissue-fraction effects. To address the challenge of accounting for partial-volume effects, and in particular, tissue-fraction effects, we propose a Bayesian approach to tissue-fraction estimation for oncological PET segmentation. Specifically, this Bayesian approach estimates the posterior mean of fractional volume that the tumor occupies within each voxel of the image. The proposed method, implemented using a deep-learning-based technique, was first evaluated using clinically realistic $2$-D simulation studies with known ground truth, in the context of segmenting the primary tumor in PET images of patients with lung cancer. The evaluation studies demonstrated that the method accurately estimated the tumor-fraction areas and significantly outperformed widely used conventional PET segmentation methods, including a U-net-based method, on the task of segmenting the tumor. In addition, the proposed method was relatively insensitive to partial-volume effects and yielded reliable tumor segmentation for different clinical-scanner configurations. The method was then evaluated using clinical images of patients with stage IIB/III non-small cell lung cancer from ACRIN $6668$/RTOG $0235$ multi-center clinical trial. Here, the results showed that the proposed method significantly outperformed all other considered methods and yielded accurate tumor segmentation on patient images with Dice similarity coefficient (DSC) of $0.82$ ($95\%$ CI: $0.78, 0.86$). In particular, the method accurately segmented relatively small tumors, yielding a high DSC of $0.77$ for the smallest segmented cross-section of $1.30~\mathrm{cm}^2$. Overall, this study demonstrates the efficacy of the proposed method to accurately segment tumors in PET images. 
\end{abstract}

\section{Introduction}
Reliable segmentation of oncological PET images is required for tasks such as PET-based radiotherapy planning and quantification of radiomic and volumetric features from PET images \citep{zaidi2009molecular,jha2017,cook2018challenges,mena201718f}. However, tumor segmentation in PET is challenging for several reasons such as partial-volume effects (PVEs), system noise, and large variabilities in the shape, texture, and location of tumors \citep{foster2014}. Tumor segmentation can be performed by having trained readers delineate the tumors manually. However, manual delineation is both labor- and time-intensive, and suffers from intra- and inter-reader variability \citep{foster2014}. To address these issues, computer-aided segmentation methods have been developed. These include methods based on thresholding, region growing, boundary detection, and stochastic modeling \citep{foster2014,SridharSUV,KassSnake,layer2015pet}. However, these methods suffer from limitations, such as requiring user inputs, sensitivity to model assumptions \citep{belhassen2010novel}, and limited ability to account for PVEs.  Learning-based methods \citep{blanc2018automatic,zhao2018tumor} have been developed to address these issues. While these methods have demonstrated promise, they typically require manual delineations for training, which are likely affected by PVEs. Thus, accounting for PVEs remains an important challenge in accurate delineation of PET images. 
\label{Sec: Introduction}

The PVEs in PET arise from two sources, namely the limited spatial resolution of PET system and the finite voxel size in the reconstructed image \citep{soret2007}. The limited spatial resolution leads to blurred tumor boundaries. The finite voxel size results in voxels containing a mixture of tumor and normal tissue. This phenomenon is referred to as tissue-fraction effects (TFEs) \citep{rousset2007partial}. A recently developed deep-learning (DL)-based technique \citep{leung2020physics} has attempted to account for PVEs arising due to the low system resolution. However, this method is not able to account for the TFEs. This shortcoming arises because this method, similar to conventional classification-based segmentation methods, is not designed or trained to model TFEs. Instead, this method is designed and trained on the task of classifying each voxel in an image as belonging to a single region. Note that while these learning-based methods can output a probabilistic measure of a voxel belonging to a region, that probability is unrelated to TFEs. Similarly, other probabilistic techniques, such as simultaneous truth and performance level estimation (STAPLE) technique \citep{dewalle2015staple}, can yield a probabilistic estimate of the true segmentation. However, again, this probabilistic estimate has no relation to TFEs. Fuzzy PET segmentation methods have attempted to account for TFEs by assigning different fuzzy levels to voxels that are partially occupied by the tumor \citep{hatt2007fuzzy,hatt2009fuzzy}. However, the goal of these methods is not to directly estimate the tumor-fraction volume within each voxel. Thus, they are not able to explicitly model TFEs.

To address the challenge of accounting for PVEs, and in particular, TFEs, while performing tumor segmentation in PET, in this manuscript, we propose a Bayesian approach to tissue-fraction estimation. Specifically, the segmentation problem is posed as a task of estimating the fractional volume that the tumor occupies within each voxel of an image. Through this strategy, we are able to explicitly model TFEs. The proposed method was developed in the context of segmenting the primary tumor in [$^{18}$F]fluorodeoxyglucose (FDG)-PET images of patients with lung cancer. 

In the next section, we develop a theoretical formalism for the proposed method. Our evaluation of the method using both clinically realistic simulations and clinical images of patients with stage IIB/III non-small cell lung cancer (NSCLC) from ACRIN $6668$/RTOG $0235$ multi-center clinical trial, is then presented in Sec.~\ref{Sec: Evaluation}, followed by the results of this evaluation, discussions, and conclusions. 

\section{Method}
\label{Sec: Method}
\subsection{Theory}
\label{Sec: Theory}
Consider a PET system imaging a radiotracer distribution, described by a vector $f(\mathbf{\rd})$, where $\mathbf{\rd} = (x,y,z)$ denotes the spatial coordinates. We denote the tracer uptake in the tumor by $f_s(\mathbf{\rd})$. The rest of the regions are referred to as background, and uptake in the background is denoted as $f_b(\mathbf{\rd})$. Thus, the tracer uptake can be represented mathematically as follows:
\begin{equation}
    f(\mathbf{\rd}) = f_b({\mathbf{\rd}}) + f_s({\mathbf{\rd}}).
    \label{eq:f_rd}
\end{equation}
We define a support function for the tumor region as $s(\mathbf{\rd})$, i.e. 
\begin{equation}
    s(\mathbf{\rd}) = \begin{cases} 
    1, & \quad \text{if} \ f_s(\mathbf{\rd}) > 0. \\
    0, & \quad \text{otherwise}.
    \end{cases}
\label{eq:s_rd}
\end{equation}
The radiotracer emits photons that are detected by the PET system, yielding projection data. Reconstruction with the projection data yields the reconstructed image, denoted by an $M$-dimensional vector $\hat{\mathbf{f}}$. Thus, the mapping from the tracer distribution to the reconstructed image is given by the operator $\Theta: \mathbb{L}_2(\mathbb{R}^3) \rightarrow \mathbb{E}^{M}$.

Denote the PET system by a linear continuous-to-discrete operator $\boldsymbol{\mathcal{{H}}}$, and let the vector $\mathbf{n}$ describe the Poisson-distributed noise. Denote the reconstruction operator, quite possibly non-linear, by $\boldsymbol{\mathcal{{R}}}$. The eventual reconstructed image is given in operator notation as follows:
\begin{equation}
    \hat{\mathbf{f}} = \boldsymbol{\mathcal{{R}}} \{\boldsymbol{\mathcal{{H}}} \mathbf{f}+ \mathbf{n}\}.
\end{equation}
In the reconstructed image, denote the volume of each voxel by $V$ and define the voxel function as $\phi_m(\mathbf{\rd})$, i.e.
\begin{equation}
    \phi_m(\mathbf{\rd}) = \begin{cases} 
    1, & \quad \text{if $\mathbf{\rd}$ lies within the $m^{\mathrm{th}}$ voxel of the PET image.} \\
    0, & \quad \text{otherwise}.
    \end{cases}
    \label{eq:phi_m}
\end{equation}
The fractional volume that the tumor occupies in the $m^\mathrm{th}$ voxel, denoted by $v_m$, is given by 
\begin{equation}
    v_m = \frac{1}{V} \int d^3 \mathfrak{r} \ s(\rd) \phi_m(\rd).
    \label{eq:v_m}
\end{equation}

Our objective is to design a method that estimates this quantity $v_m$ from the reconstructed image $\hat{\mathbf{f}}$ for all $M$ voxels. Denote the estimate of $v_m$ by $\hat{v}_m$. Further, denote the $M$-dimensional vector ${\{v_1, v_2, \ldots, v_M\}}$ by $\mathbf{v}$, and denote the estimate of $\mathbf{v}$ by $\hat{\mathbf{v}}$.

Estimating $\mathbf{v}$ from the reconstructed image is an ill-posed problem due to the null spaces of the $\boldsymbol{\mathcal{{H}}}$ and $\boldsymbol{\mathcal{{R}}}$ operator. Thus, we take a Bayesian approach to estimate $\hat{\mathbf{v}}$. We first need to define a cost function that penalizes deviation of $\mathbf{v}$ from $\hat{\mathbf{v}}$. A common cost function is the ensemble mean squared error (EMSE), which is the mean squared error averaged over noise realizations and the true values $\mathbf{v}$. However, in our case, the variable $\hat{v}_m$ is constrained to lie within $\left[0,1\right]$, and the EMSE loss does not directly incorporate this constraint. In contrast, using the binary cross-entropy (BCE) loss as the penalizer allows us to incorporate this constraint on $\hat{v}_m$ directly, as also suggested in \cite{creswell2017denoising}. The BCE loss between $v_m$ and $\hat{v}_m$, denoted by $l_{BCE}(v_m, \hat{v}_m)$, is given by
\begin{equation}
    l_{BCE}(v_m, \hat{v}_m) =  -v_m\mathrm{log}(\hat{v}_m) - (1-v_m) \mathrm{log}(1-\hat{v}_m).
    \label{eq: BCE loss}
\end{equation}

We define our cost function $C(\mathbf{v}, \hat{\mathbf{v}})$ as the negative of aggregate BCE loss over all voxels averaged over the joint distribution of true values $\mathbf{v}$ and noise realizations $\hat{\mathbf{f}}$. The cost function is then given by
\begin{equation}
\begin{split}
    C(\mathbf{v},\hat{\mathbf{v}}) 
    &= - \int d^M \mathbf{\hat{f}} \int d^M \mathbf{v} \ \mathrm{pr}(\mathbf{\hat{f}},\mathbf{v}) \sum_{m=1}^M l_{BCE}(v_m, \hat{v}_m) \\
    &= - \int d^M \mathbf{\hat{f}} \ \mathrm{pr}(\mathbf{\hat{f}}) \int d^M \mathbf{v} \ \mathrm{pr}(\mathbf{v}|\mathbf{\hat{f}}) \sum_{m=1}^M l_{BCE}(v_m, \hat{v}_m),
\end{split}
\label{eq: Cost function before sim}
\end{equation}
where in the second step we have expanded $\mathrm{pr}(\mathbf{\hat{f}},\mathbf{v})$ using the conditional probability. Inserting Eq. \eqref{eq: BCE loss} into Eq. \eqref{eq: Cost function before sim}, we obtain
\begin{equation}
    C(\mathbf{v},\hat{\mathbf{v}}) = \int d^M \mathbf{\hat{f}} \ \mathrm{pr}(\mathbf{\hat{f}}) \int d^M \mathbf{v} \ \mathrm{pr}(\mathbf{v}|\mathbf{\hat{f}}) \left[ \sum_{m=1}^M v_m\mathrm{log}(\hat{v}_m) + (1-v_m) \mathrm{log}(1-\hat{v}_m) \right].
    \label{eq: Cost function final form}
\end{equation}

To estimate the point at which this cost function is minimized, we differentiate the cost function with respect to the vector $\hat{\mathbf{v}}$ and set that equal to zero. Because $\mathrm{pr}(\hat{\mathbf{f}})$ is always nonnegative, the cost function is minimized by setting the derivative of inner integral in Eq. \eqref{eq: Cost function final form} equal to zero, i.e.
\begin{equation}
    \frac{\partial}{\partial \hat{\mathbf{v}}} \int d^M \mathbf{v} \ \mathrm{pr}(\mathbf{v}|\mathbf{\hat{f}}) \left[ \sum_{m=1}^M v_m\mathrm{log}(\hat{v}_m) + (1-v_m) \mathrm{log}(1-\hat{v}_m) \right] = 0.
\label{eq: derivative of inner integral of Cost function}
\end{equation}
This is then equivalent to performing component-wise differentiation and setting each differentiated component to $0$ \citep{barrett2013foundations}. For the $m^\mathrm{th}$ component of Eq. \eqref{eq: derivative of inner integral of Cost function}, we get
\begin{equation}
\begin{split}
    & \frac{\partial}{\partial \hat{v}_m} \int dv_m \ \mathrm{pr}(v_m|\mathbf{\hat{f}}) \left[ v_m\mathrm{log}(\hat{v}_m) + (1-v_m) \mathrm{log}(1-\hat{v}_m) \right] \\
    &= \frac{\partial}{\partial \hat{v}_m} \int dv_m \ \mathrm{pr}(v_m|\mathbf{\hat{f}}) \left[ v_m \left\{ \mathrm{log}(\hat{v}_m) - \mathrm{log}(1-\hat{v}_m) \right\} + \mathrm{log}(1-\hat{v}_m) \right] \\
    &= 0.
\end{split}
\label{eq: derivative of mth component of cost function}
\end{equation}
Since $\int dv_m \ \mathrm{pr}(v_m|\mathbf{\hat{f}}) = 1$,
the solution to Eq. \eqref{eq: derivative of mth component of cost function}, denoted by $\hat{v}_m^*$, is given by
\begin{equation}
    \hat{v}_m^* = \int dv_m \ \mathrm{pr}(v_m|\mathbf{\hat{f}}) v_m.
    \label{eq: vhatm*}
\end{equation}
Equivalently, in vector notation, we get
\begin{equation}
    \hat{\mathbf{v}}^* = \int d^M\mathbf{v} \ \mathrm{pr}(\mathbf{v}|\mathbf{\hat{f}}) \mathbf{v},
    \label{eq: vhat*}
\end{equation}
which is simply the posterior-mean estimate of $\mathbf{v}$. Note that the same estimator is obtained when the cost function is the EMSE between $\mathbf{v}$ and $\hat{\mathbf{v}}$. Thus, by minimizing the cost function in Eq. \eqref{eq: Cost function final form}, we obtain an optimal estimator that achieves the lowest mean squared error among all possible estimators. We can further show that this estimator is unbiased in a Bayesian sense (proof provided in \ref{appendix b}). 

In summary, we have shown that by developing an optimization procedure that minimizes the cost function defined in {Eq.~\eqref{eq: Cost function final form}}, we obtain a posterior-mean estimate of the tumor-fraction volumes in each voxel of the reconstructed image. This estimator yields the lowest mean squared error among all possible estimators. Further, this estimator is unbiased in a Bayesian sense.

\subsection{Implementation of the proposed technique}
\label{Sec: Implementation of the estimator}
While we have developed the theoretical formalism in $3$-D, in this manuscript, the method was implemented and evaluated on a simplified per-slice basis. Thus, for each pixel in the $2$-D reconstructed image, the optimizer was designed to yield the posterior mean estimate $\hat{\mathbf{a}}^*$ of the true tumor-fraction area (TFA), which we denote by $\mathbf{a}$. We now describe the procedure to implement this optimizer.

\begin{figure}[h]
    \centering
    \includegraphics[width=\textwidth]{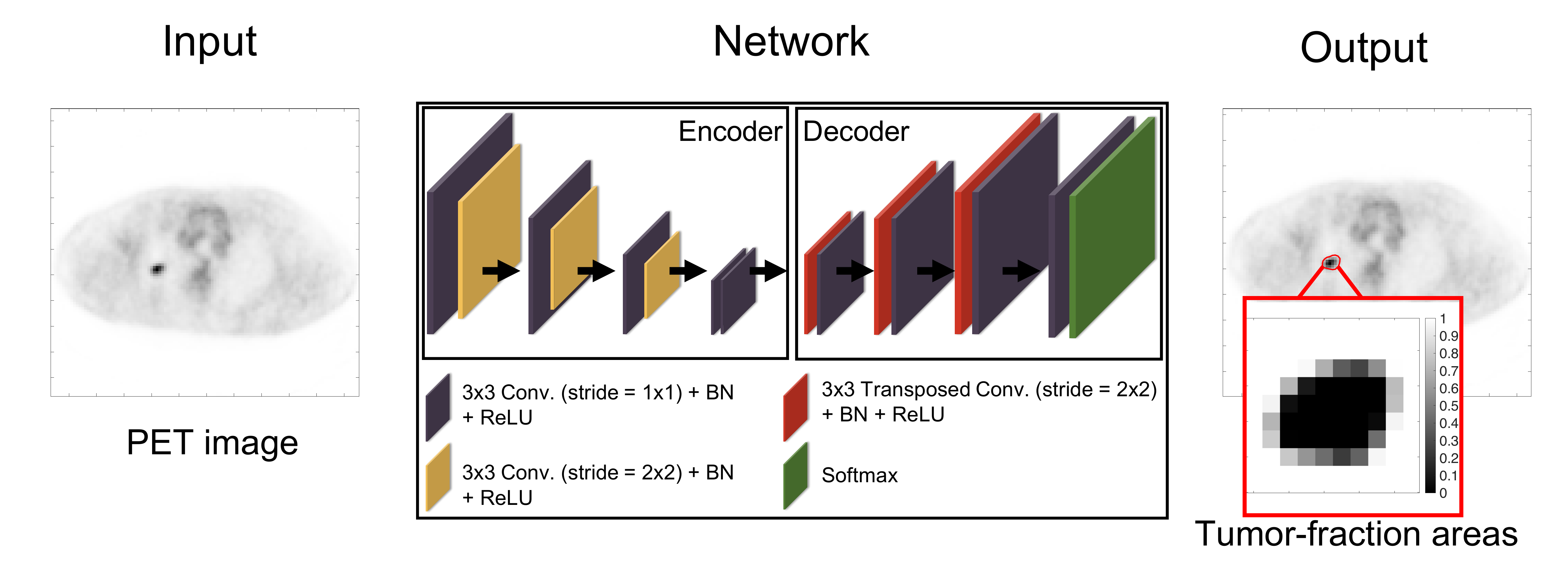}
    \caption{Illustration of the developed optimization procedure by constructing an encoder-decoder network. Conv.: convolutional layer; BN: batch normalization; ReLU: rectified linear unit.}
    \label{fig:Network_architecture}
    \captionsetup{justification=centering}
\end{figure}

Estimating the posterior mean $\hat{\mathbf{a}}^*$ requires sampling from the posterior distribution $\mathrm{pr}(\mathbf{a}|\hat{\mathbf{f}})$. Sampling from this distribution is challenging as this distribution is high-dimensional and does not have a known analytical form. To address this issue, the proposed method was implemented using a supervised learning-based approach. Specifically, an encoder-decoder network was constructed, as shown in Fig.~\ref{fig:Network_architecture}. During the training phase, this network is provided with a population of PET images, and the corresponding ground-truth TFA map, i.e. the vector $\mathbf{a}$ for each image, as described in Sec. \ref{Sec: Theory}. The network, by minimizing the cost function defined in Eq.~\eqref{eq: Cost function final form} over this population of images, becomes trained to yield the posterior-mean estimate of $\mathbf{a}$ given the input PET image. 

The network architecture is similar to those for estimation tasks, such as image denoising \citep{creswell2017denoising} and image reconstruction \citep{nath2020deep}. To summarize, the network is partitioned into a contracting and an expansive path. The contracting path learns the spatial information from the input PET images and the expansive path maps the learned information to the estimated TFA map for each input image. Skip connections with element-wise addition were applied to feed the features extracted in the contracting path into the expansive path to stabilize the training and improve the learning performance \citep{mao2016image}. In the final layer, the network yields the estimate of the TFAs. A detailed description of the network architecture is provided in \ref{appendix a} (Table~\ref{tab:network architecture}). 

As outlined in Sec. \ref{Sec: Introduction}, the goal of the proposed method is to explicitly model the TFEs while performing tumor segmentation. Our training strategy and network architecture are specifically designed for this goal by defining the ground truth as the TFAs for each image. We contrast this to the conventional DL-based segmentation methods, where, in the ground truth, each pixel is exclusively assigned to the tumor or the normal tissue class and the network is trained to classify each pixel as either tumor or background. Further, as mentioned above, while the conventional DL-based methods can output a probabilistic estimate for each image pixel, this estimate is only a measure of classification uncertainty, and thus has no relation to TFEs, unlike the proposed method.

The network was trained via the Adam optimization algorithm \citep{kingma2014adam}. In the various experiments mentioned later, the network hyperparameters were optimized on a training set via five-fold cross validation. The network training was implemented in Python 3.6.9, Tensorflow 1.14.0, and Keras 2.2.4. Experiments were performed on a Linux operating system with two NVIDIA Titan RTX graphics processing unit cards.

\section{Evaluation} 
\label{Sec: Evaluation}
Evaluating the proposed method requires access to ground truth where either the ground-truth TFA map or a surrogate for the true TFA map, such as tumor delineations defined by trained readers, are known. In Sec.~\ref{Sec: evaluation using clinically realistic simulation}, we first evaluated the proposed method using clinically realistic simulation studies, where the ground-truth TFA map was known. In these studies, the support of tumor can be described at a very high resolution, simulating~$s({\mathbf{\rd}})$ in Eq.~\eqref{eq:s_rd}. From this high-resolution description, the true TFA within each image pixel can be computed using Eq.~\eqref{eq:v_m}, thus providing the TFA map. Realistic simulation studies also model imaging physics and variability in patient populations. Thus, these studies provide a rigorous mechanism to evaluate the method. However, we recognize that simulation studies may have limitations in modeling all aspects of system instrumentation, patient physiology, and patient-population variability, especially in multi-center settings, accurately. Thus, it is important to assess the performance of the method using patient studies, ideally with multi-center trial data. For this purpose, in Sec.~\ref{Sec: evaluation using patient studies}, we evaluated the proposed method on clinical images from the ACRIN $6668$/RTOG $0235$ multi-center clinical trial, where trained-reader-defined segmentations were used as the surrogate ground truth. We first describe the performance metrics used to quantitatively evaluate the proposed method.

\subsection{Evaluation metrics}
\label{sec: evaluation metrics}
Since the proposed method is an estimation-based segmentation approach, our evaluation used performance metrics for both the task of estimating the true TFA map and of segmenting the tumor.

\subsubsection{Evaluation on estimation performance}
Performance on the estimation task was evaluated using the EMSE between the true and estimated TFA maps. EMSE provides a combined measure of bias and variance over the distribution of true values and noise realizations, and is thus considered as a comprehensive figure of merit for estimation tasks \citep{barrett2013foundations}. The error in the estimate of the TFA maps and the tumor area was quantified using the pixel-wise EMSE and normalized area EMSE, respectively. Denote $\langle \ldots \rangle_X$ as the expected value of the quantity in the brackets when averaged over the random variable $X$. The pixel-wise EMSE is given by
\begin{equation}
    \mathrm{Pixel \mhyphen wise \ EMSE} = \left \langle \left \langle ||\hat{\mathbf{a}}-\mathbf{a}||_2^2 \right \rangle_{\hat{\mathbf{f}}|\mathbf{a}} \right \rangle_\mathbf{a}.
\end{equation}
The normalized area EMSE denotes the EMSE between the true and estimated areas of each tumor, normalized by the true areas. The true and estimated areas, denoted by $A$ and $\hat{A}$, are given by the $\mathbb{L}_1$ norms of $\mathbf{a}$ and $\hat{\mathbf{a}}$, respectively. The normalized area EMSE is then given by
\begin{equation}
    \mathrm{Normalized \ area \ EMSE} = \left \langle \left \langle \frac{|\hat{A}-A|^2}{A^2} \right \rangle_{\hat{\mathbf{f}}|A} \right \rangle_A.
\end{equation}

We have shown (Eq.~\eqref{eq: unbiased in Bayesian sense} in \ref{appendix b}) that the proposed method yields an unbiased estimate of $\mathbf{a}$ in a Bayesian sense. To verify this, the ensemble-average bias was computed. This term, denoted by $\overline{\mathbf{b}}$, is an $M$-dimensional vector ${\{\overline{b}_1, \overline{b}_2, \ldots, \overline{b}_M\}}$, with the $m^{\mathrm{th}}$ element of the vector quantifying the average bias of the estimated TFA within the $m^{\mathrm{th}}$ pixel. Consider a total of $P$ tumor images and $N$ noise realizations for each tumor image. Let $a_{mnp}$ and $\hat{a}_{mnp}$ denote the true and estimated TFA within the $m^{\mathrm{th}}$ pixel for the $n^{\mathrm{th}}$ noise realization in the $p^{\mathrm{th}}$ tumor image. The $m^{\mathrm{th}}$ component of ensemble-average bias, $\overline{b}_m$, is then given by \begin{equation}
\overline{b}_m = \frac{1}{P} \sum_{p=1}^P \frac{1}{N} \sum_{n=1}^N \ \left[\hat{a}_{mnp} - a_{mnp}\right].
\end{equation} 
The proximity of the elements of $\overline{\mathbf{b}}$ to $0$ would indicate that the estimator was unbiased in a Bayesian sense.

\subsubsection{Evaluation on segmentation performance}
\label{sec:Evaluation on segmentation performance}
The proposed method estimates the TFA within each pixel, which is a continuous-valued output. For evaluation of segmentation methods that yield such non-binary output, as in \cite{taha2015metrics}, the spatial-overlap-based metrics can be derived based on the four cardinalities of confusion matrix, namely the true positives (TP), false positives (FP), true negatives (TN), and false negatives (FN). The four cardinalities are given by
\begin{equation}
\begin{split}
    &\mathrm{TP} = \sum_{m=1}^{M} \mathrm{min} \left( \hat{a}_m,a_m \right) \ \ \ \ \ \ \ \ \ \ \ \ \ \mathrm{FP} = \sum_{m=1}^{M} \mathrm{max} \left( \hat{a}_m-a_m, 0 \right) \\
    &\mathrm{TN} = \sum_{m=1}^{M} \mathrm{min} \left( 1 - \hat{a}_m, 1 - a_m \right) \ \ \  \mathrm{FN} = \sum_{m=1}^{M} \mathrm{max} \left( a_m - \hat{a}_m, 0 \right).
\end{split}{}
\end{equation}
The spatial-overlap metric of Dice similarity coefficient (DSC) and Jaccard similarity coefficient (JSC) were used to measure the agreement between the true and estimated segmentation. The DSC and JSC are defined as 
\begin{equation}
    \mathrm{DSC} = \frac{2\mathrm{TP}}{2\mathrm{TP} + \mathrm{FP} + \mathrm{FN}}, \ \ \  \ \ \ \mathrm{JSC} = \frac{\mathrm{TP}}{\mathrm{TP + FP + FN}}.
\end{equation}
Higher values of DSC and JSC indicate higher segmentation accuracy. These metrics were reported as mean values with 95\% confidence intervals (CIs). Statistical significance was assessed via a paired sample \textit{t}-test, with a \textit{p}-value $< 0.01$ inferring statistically significant difference. 

We also qualitatively evaluated the performance of the proposed method on the task of estimating the TFA map. For this purpose, ground-truth and estimated tumor topographic maps were first constructed from the true and estimated TFA maps using the contour function in MATLAB (MathWorks, Natick, Mass). Specifically, the tumor topographic map shows the topography of the TFA map by means of isocontours. Then, isocontours corresponding to the true and estimated TFA maps were plotted for the TFA values of $0$, $1/3$, $2/3$, and $1$. A TFA of 0 implies that no area within that pixel contains the tumor, while a TFA of 1 implies that the entire pixel area is the tumor.

\subsection{Evaluation of the proposed method using clinically realistic simulation studies}
\label{Sec: evaluation using clinically realistic simulation}
This evaluation study was conducted in the context of segmenting the primary tumor in FDG-PET images of patients with lung cancer. The study quantitatively evaluated the accuracy of the method, compared the method to existing techniques, studied the sensitivity of the method to PVEs, and also studied the performance of the method for different clinical-scanner configurations. In each evaluation, clinically realistic simulated PET images with known ground-truth tumor properties were generated, as described in Sec.~\ref{Sec: generating simulated images}. The generated data was split into training and test sets. The proposed method was trained and cross-validated using the training set. The performance of the method was then evaluated using the independent test set. The evaluation study used clinical images, was retrospective, IRB-approved, and HIPAA-compliant
with a waiver of informed consent.

\subsubsection{Generating realistic simulated PET images}
\label{Sec: generating simulated images}
The simulation strategy advances on a previously proposed approach to simulate PET images \citep{leung2020physics}. Briefly, in the first step, realistic tumor-tracer distribution was simulated at a very high resolution, so that the simulated tumor can be described potentially as a continuous object, equivalent to $f_s(\mathbf{r})$ in Eq.~\eqref{eq:f_rd}, except that $\mathbf{r} = (x,y)$ is a $2$-D vector. Specifically, the pixel size in the simulated tumor image was $0.13~\mathrm{mm}$. This was $1/32$ of the resolution in the patient image. The shapes, sizes, and intensities of simulated tumors were sampled from the corresponding distribution derived from clinical images, so that the simulated tumors had variabilities as observed in patient populations. An advancement on the approach proposed in \cite{leung2020physics} was to simulate intra-tumor heterogeneity using a stochastic lumpy object model \citep{rolland1992effect}. Existing clinical PET images containing the lung region but with no tumor present were selected as templates to ensure tumor-background realism and account for inter-patient variability. The projection data for the simulated tumor and background were generated using a PET simulation software \citep{leung2020physics}. Since the simulated tumor had higher resolution compared to the background, we had different projection models for the tumor and background separately. The projection data for the tumor and background were then added, enabling the impact of image reconstruction on the tumor appearance and noise texture to be inherently incorporated \citep{Ma2016}. Reconstruction was performed using a $2$-D ordered subset expectation maximization (OSEM) algorithm. We have validated the realism of the images simulated using this approach \citep{liu2021evaluation}. Detailed simulation and reconstruction parameters will be provided for each of the studies mentioned below. 

\subsubsection{Evaluating accuracy of the proposed method and comparing to other segmentation methods}
\label{Sec: Evaluating accuracy and comparing to other segmentation methods}
We quantitatively compared the proposed method to several commonly used semi-automated PET segmentation methods, including $40$\% SUV-max thresholding  \citep{SridharSUV}, active-contour-based Snakes  \citep{KassSnake}, and Markov random fields-Gaussian mixture model (MRF-GMM) \citep{layer2015pet,JhaGMM}. The method was also compared to a fuzzy segmentation method, namely the fuzzy local information C-Means clustering algorithm (FLICM) \citep{krinidis2010robust}. Further, the method was compared to a U-net-based PET segmentation method \citep{leung2020physics}. The ground truth for training this U-net-based method was defined such that each voxel was classified as either tumor or background. For all the semi-automated segmentation methods, the tumor location was provided by manually generating a rectangular region of interest (ROI) containing the tumor. In contrast, the proposed and U-net-based method did not require any manual input and were fully automated.

To generate the simulated images for this study, following the procedure in Sec. \ref{Sec: generating simulated images}, we used $318$ $2$-D slices from $32$ patients for the background portion of the image. The simulated PET system had a spatial resolution of $5$ mm full width at half maximum (FWHM). The projection data were reconstructed using the OSEM algorithm with $21$ subsets and $2$ iterations, similar to the PET reconstruction protocol for the patient images. The reconstructed pixel size was $4.07~\mathrm{mm} \times 4.07~\mathrm{mm}$. The network was trained and cross-validated using $9,540$  images with $5$-fold cross validation. Evaluation was then performed on $2,070$ completely independent test images, which were generated using $69$ $2$-D slices from $7$ patients.

\subsubsection{Evaluating sensitivity of the proposed method to PVEs} 
\label{Sec: Evaluating sensitivity of the proposed method to PVEs}
To conduct this evaluation, similar to \cite{le2011evaluation} and \cite{leung2020physics}, we studied the performance of the method as a function of tumor area. For this purpose, all test images were grouped based on the range of the true tumor area. Specifically, the tumor areas were binned with a bin width of 2 $\mathrm{cm}^2$. For each test image, PVEs-affected tumor masks were generated by applying a rectangular filter to the ground-truth tumor mask, following the strategy in \cite{leung2020physics}. This filter modeled the resolution degradation due to the forward projection and the reconstruction process. The tumor area measured using the proposed method and the PVEs-affected tumor area in all the test images were obtained and divided by the true tumor area. A ratio of unity would indicate that the output was insensitive to PVEs. A ratio lower or higher than unity would indicate an underestimation or overestimation of the true tumor area, respectively, showing that the segmentation output was affected by PVEs \citep{de2009lesion}. 

\subsubsection{Evaluating accuracy of the proposed method for different clinical-scanner configurations} 
\label{clinical configuration}
For this purpose, we simulated two PET systems with configurations similar to the Siemens Biograph 40 and Biograph Vision scanners. The PET images reconstructed from these two scanners had different pixel sizes, as dictated by the protocol. The Biograph 40 generated images of $128 \times 128$ pixels, while the Biograph Vision generated images of $192 \times 192$ pixels. Details of the PET scanner acquisition and reconstruction parameters are provided in \ref{appendix a} (Table \ref{tab:TFE_PET_specs}). Clinical PET images of patients with lung cancer were obtained from these scanners. Using these clinical scans and following the simulation procedure described in Sec.~\ref{Sec: generating simulated images}, a total of $5,520$ and $6,120$ simulated PET images were generated for each scanner, respectively. These were used for optimizing and training the network. Next, the trained network was tested on $1,200$ and $1,320$ independent simulated images, respectively. The performance of the proposed method was also compared to the U-net-based method.

\subsection{Evaluation of the proposed method using clinical multi-center PET images}
\label{Sec: evaluation using patient studies}
We next evaluated the proposed method using clinical PET images. For this purpose, we used de-identified patient data from the ACRIN $6668$/RTOG $0235$ multi-center clinical trial \citep{machtay2013prediction,kinahanTCIAdata} available from The Cancer Imaging Archive \citep{clark2013cancer}. In this evaluation study, FDG-PET images of $78$ patients with inoperable stage IIB/III NSCLC were included. Detailed patient demographics with clinical characteristics are provided in \ref{appendix a} (Table~\ref{tab:patient_demographics}). As in \cite{machtay2013prediction}, the standard imaging protocol involved recommended dose level from $10$ to $20$ mCi and image acquisition beginning $50$ to $70$ minutes after FDG injection. PET images were acquired from ACRIN-qualified clinical scanners \citep{scheuermann2009qualification}, with attenuation, scatter, random, normalization, decay, and deadtime correction applied in the reconstruction protocol. For all the $78$ patients, the PET images were of size $128 \times 128$, with the pixel size ranging from $4.69~\mathrm{mm}$ to $5.47~\mathrm{mm}$. Detailed reconstruction parameters are provided in \ref{appendix a} (Table~\ref{tab:scanner_info}).

Evaluation of the proposed method would require the knowledge of true TFA maps. For this purpose, a board-certified nuclear-medicine physician (J.C.M) with more than $10$ years of experience in reading PET scans identified the primary tumor of each patient by reviewing the PET, CT, and fused PET/CT images along axial, sagittal, and coronal planes using MIM Encore (MIM Software, version 6.9.3). The radiologist was asked to delineate a continuous (un-pixelated) boundary for each identified tumor. For each tumor, the radiologist drew an external tumor boundary and considered the whole volume within that boundary as belonging to the tumor class. A workflow was created in MIM to assist the radiologist with this delineation task. The radiologist used a MIM-based edge-detection tool to segment the tumor in 3-D on the fused PET/CT image, by placing the cursor at the center of the tumor and dragging it out until the three orthogonal guiding lines reached the tumor boundary. The radiologist then examined and adjusted the segmentation to make it more accurate and also account for PVEs. This manual segmentation was continuous and allowed for a voxel to consist of a mixture of tumor and normal tissues. The segmentation was saved at a higher resolution than that of the PET image.

From this manual segmentation, we obtained a discrete version of the tumor mask, $s(\mathbf{r})$, as defined in Eq.~\eqref{eq:s_rd}, for each $2$-D PET slice and at a higher resolution than the PET image. Specifically, the pixel size in the tumor mask was 1/8 of that in the PET image. This resolution was chosen since more fine sampling did not cause changes in the definition of the tumor mask. Let this high-resolution manual segmentation be an $N$-dimensional vector ($N$~$>$~$M$), where we recall that $M$ was the dimension of the PET image. Denote the pixel function in this high-resolution space by $\phi_n^{manual} (\mathbf{r})$, following the similar definition in Eq.~\eqref{eq:phi_m}. Define an $N$-dimensional vector $\psi(\mathbf{r})$ with each element of this vector defined as
\begin{equation}
\psi_n(\mathbf{r}) = 
    \begin{cases}
        1 & \parbox[t]{.6\textwidth}{if pixel \textit{n} in the manual segmentation is assigned to tumor class.}\\
        0 & \text{otherwise.}
    \end{cases}
\end{equation}
Denote the pixel area of the PET image by $A$. We computed the ground-truth TFA within each image pixel as follows:
\begin{equation}
    a_m = \frac{1}{A} \sum_{n=1}^N \psi_n(\mathbf{r}) \int d^2\mathbf{r} \ \phi_n^{manual} (\mathbf{r})\phi_m(\mathbf{r}),
\end{equation}
where the integral computes fractional area that $n^\mathrm{th}$ pixel in the manual segmentation occupies within the $m^\mathrm{th}$ pixel of the PET image. The network was then trained to estimate the posterior mean of $a_m$ for the $m^\mathrm{th}$ image pixel, following the training strategy described in Sec.~\ref{Sec: Implementation of the estimator}.

The network was trained and cross-validated using $565$ $2$-D slices from $61$ out of $78$ patients. The trained network was then evaluated on $140$ completely independent $2$-D slices from the remaining $17$ patients. The performance of the proposed method was compared to the other segmentation methods, described in Sec.~\ref{Sec: Evaluating accuracy and comparing to other segmentation methods}, both quantitatively and qualitatively, using the procedure and metrics described in Sec.~\ref{sec:Evaluation on segmentation performance}. 

\section{Results}
\label{Sec: results}

\subsection{Evaluation of the proposed method using clinically realistic simulation studies}

\subsubsection{Evaluating accuracy of the proposed method and comparing to other segmentation methods} \label{Sec: results of Evaluating accuracy}

Quantitatively, the proposed method significantly outperformed ($p~<~ 0.01$) all other considered methods, including the U-net-based method, on the basis of the pixel-wise EMSE, normalized area EMSE, DSC, and JSC (Fig.~\ref{fig:simul_metrics}, Table~\ref{tab:simulation_metric_table} in \ref{appendix a}). The proposed method yielded the lowest pixel-wise EMSE, the lowest normalized area EMSE of $0.02$, the highest DSC of $0.90$ ($95$\% CI: $0.90, 0.91$), and the highest JSC of $0.83$ ($95$\% CI: $0.83, 0.84$). In addition, all the elements of the ensemble-average bias map were close to $0$, providing the evidence that the method yielded an unbiased Bayesian estimate of the TFA map, as shown in Sec.~\ref{Sec: Theory}. Further, the proposed method accurately segmented relatively small tumors, and in particular, yielded high DSC of $0.84$ for the smallest segmented tumor axial cross-section of $0.88~\mathrm{cm}^2$ in area. The diameter of this tumor was approximately twice the FHWM of the system resolution.

\begin{figure}[h]
    \centering
    \includegraphics[width=\textwidth]{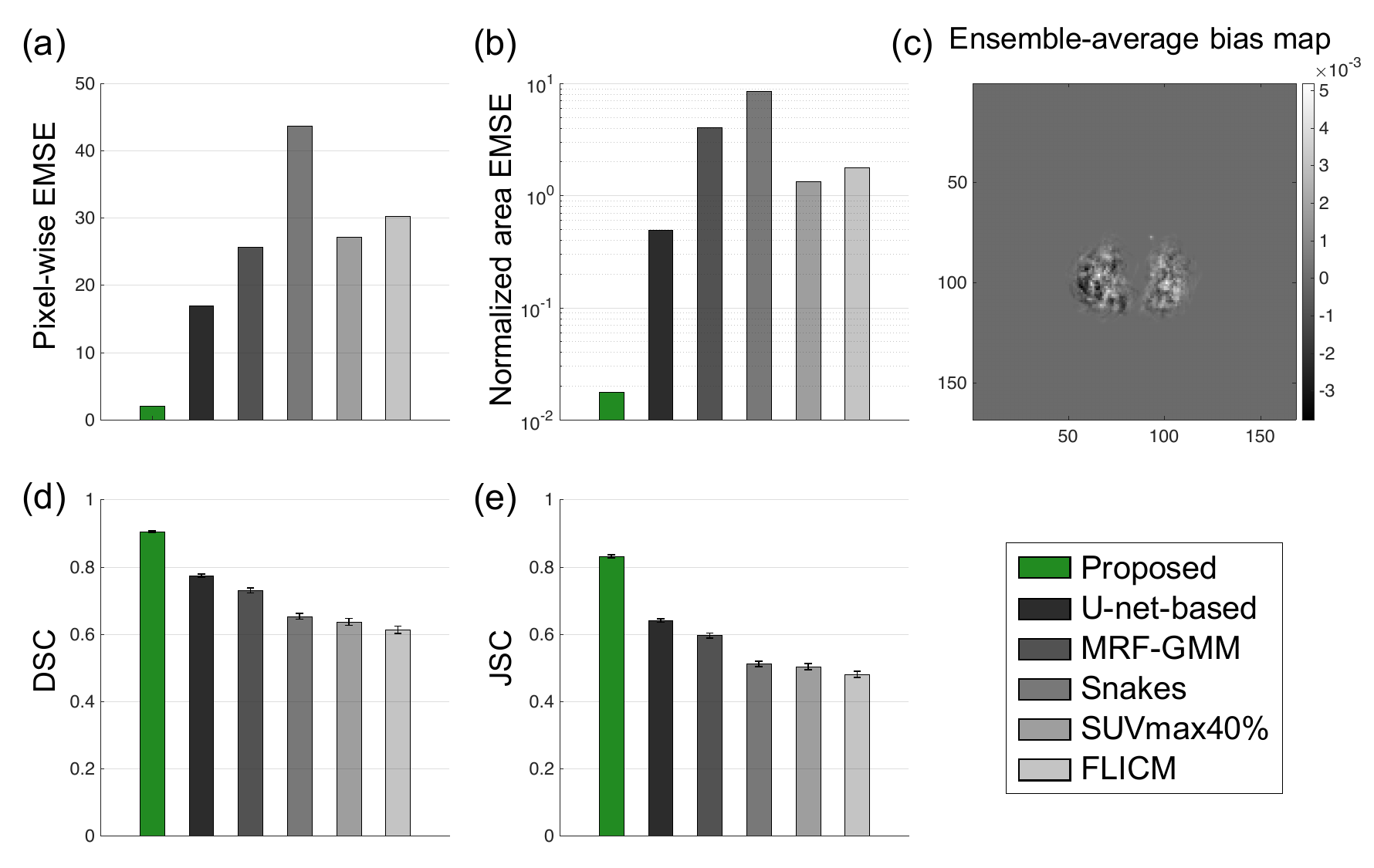}
    \caption{Evaluation result using clinically realistic simulation studies: (a) the pixel-wise EMSE between the true and estimated tumor-fraction areas; (b) the normalized area EMSE between the measured and true tumor areas (plot displayed in log scale on y-axis for better visualization); (c) the ensemble-average bias of the proposed method; the (d) Dice similarity coefficient and (e) Jaccard similarity coefficient between the true and estimated segmentations.}
    \label{fig:simul_metrics}
    \captionsetup{justification=centering}
\end{figure}

\begin{figure}[h]
    \centering
    \includegraphics[width=\textwidth]{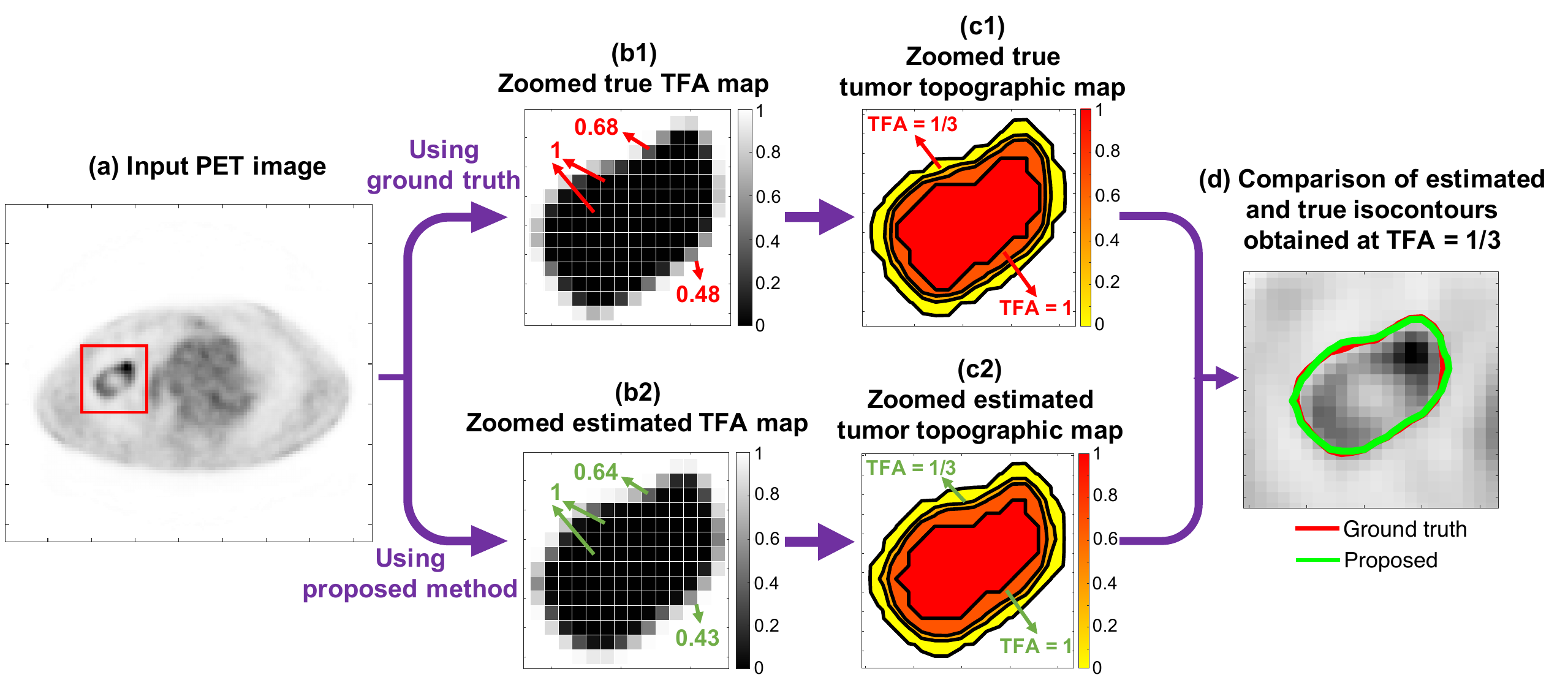}
    \caption{Illustration of the procedure to obtain isocontours from the ground-truth TFA map and the TFA map estimated by the proposed method.}
    \label{fig:isocontour_diagram}
\end{figure}

\begin{figure}[h]
    \centering
    \includegraphics[width=\textwidth]{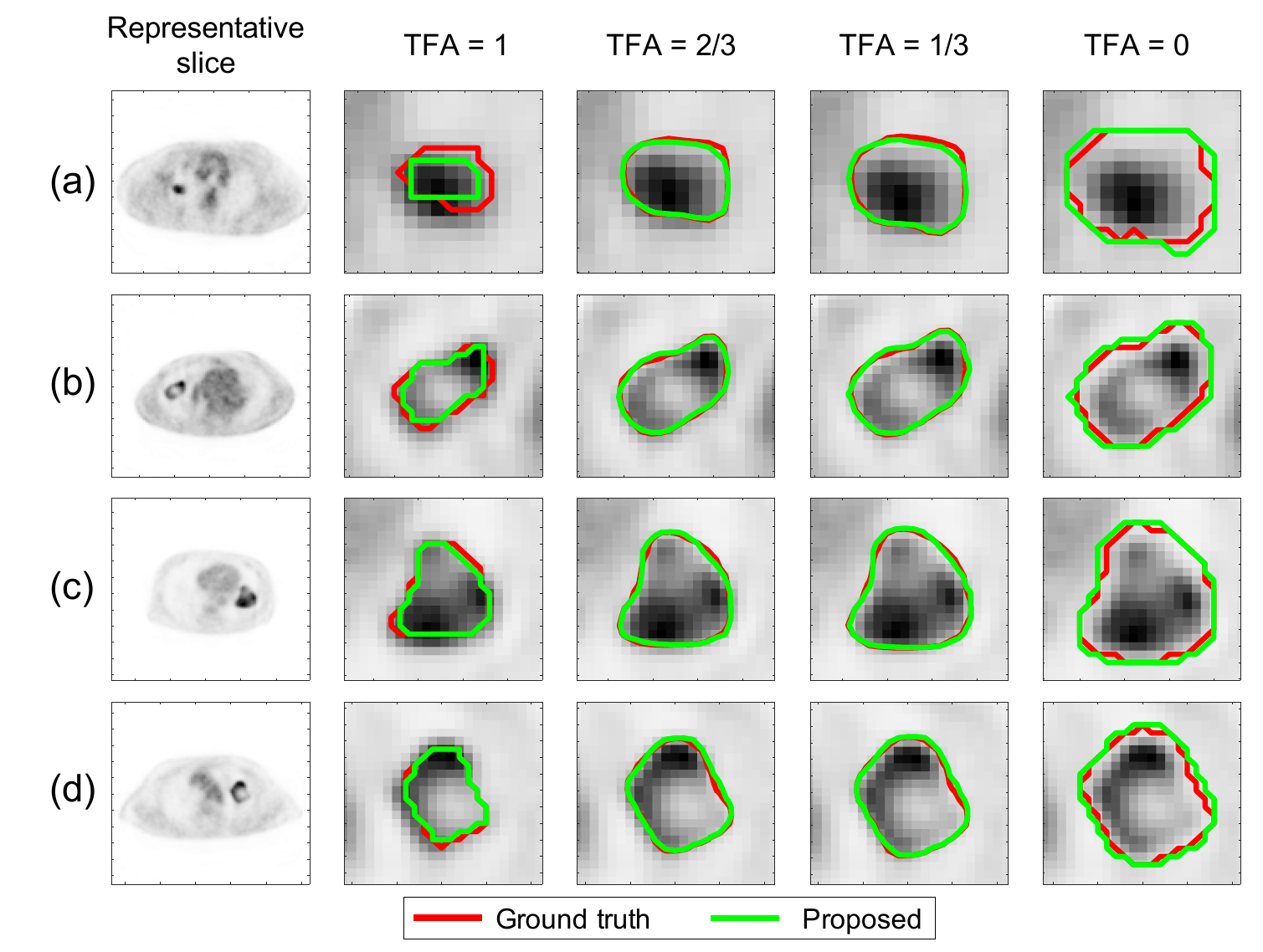}
    \caption{Evaluation result using clinically realistic simulation studies: comparison between the estimated isocontours using the proposed method (green) and the ground-truth isocontours (red), defined from set of points at four TFA values (0, 1/3, 2/3, 1).}
    \label{fig:simul_contour}
\end{figure}

We next qualitatively show the performance of the proposed method on the task of estimating the TFA map, following the procedure described in Sec.~\ref{sec:Evaluation on segmentation performance}. We first illustrate the procedure to obtain the isocontours from the ground-truth and estimated TFA maps for a representative tumor (Fig.~\ref{fig:isocontour_diagram}). We then followed this procedure to obtain the isocontours from the TFA maps for different cases. In Fig. 4, the comparisons between the true and estimated isocontours for representative slices at four different TFA values are shown. We observe that the proposed method yielded isocontours close to the true isocontours at different considered TFA values. In addition, the method yielded accurate segmentation for different tumor types, including those with substantial intra-tumor heterogeneity as best observed in Fig.~\ref{fig:simul_contour}(b-d).

\subsubsection{Evaluating sensitivity of the proposed method to PVEs}
\begin{figure}[h]
    \centering
    \includegraphics[width=\textwidth]{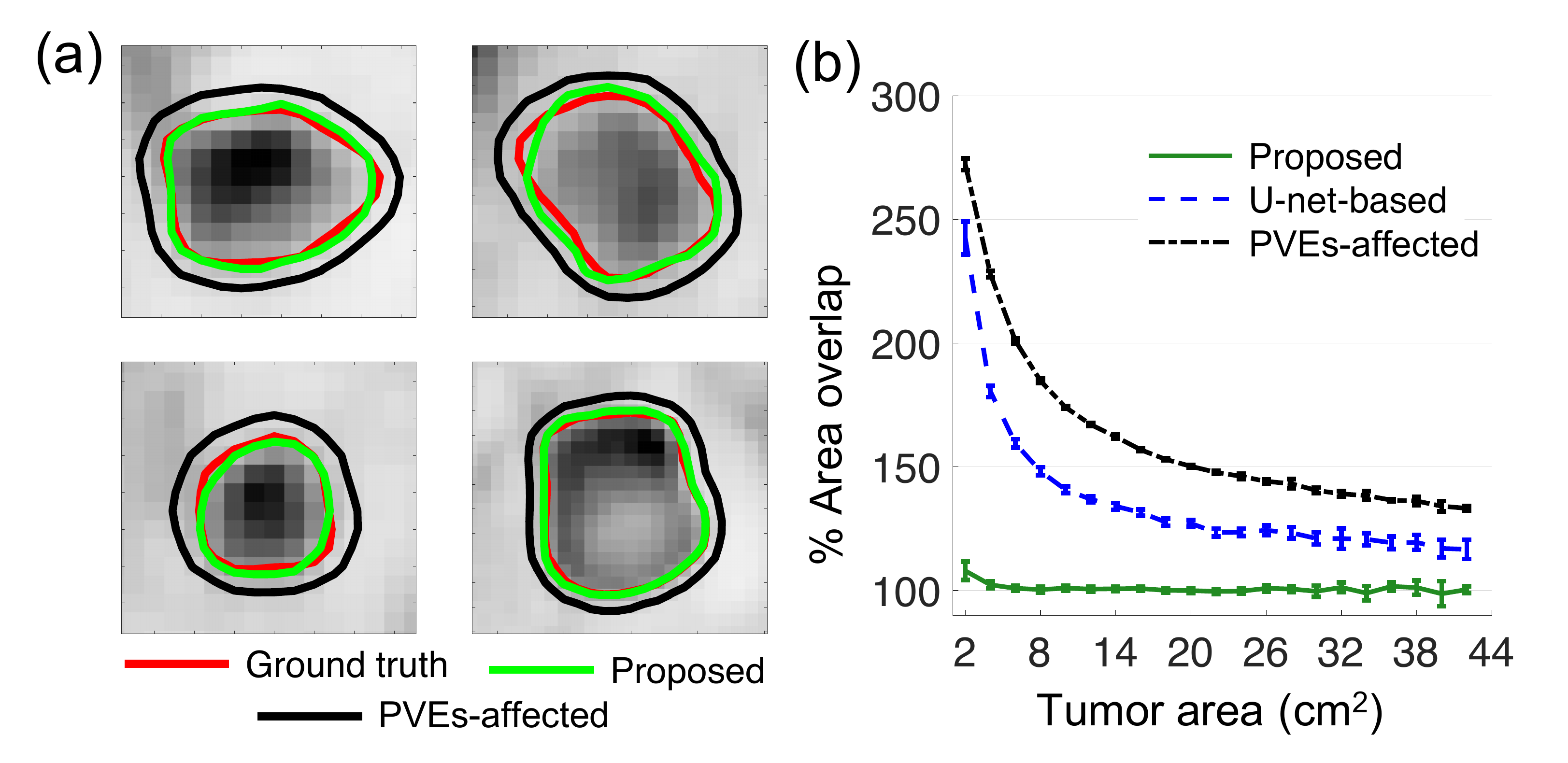}
    \caption{Evaluation result using clinically realistic simulation studies: (a) qualitative comparison between the isocontours generated from the PVEs-affected TFA maps and the isocontours generated from the estimated TFA maps using the proposed method. The isocontours were defined as the set of points with TFA equal to 0.5. (b) quantitative evaluation of the sensitivity of the proposed method to PVEs. Results obtained using the U-net-based method are also shown.}
    \label{fig:simul_pve_exp}
\end{figure}

\begin{figure}[h]
    \centering
    \includegraphics[width=\textwidth]{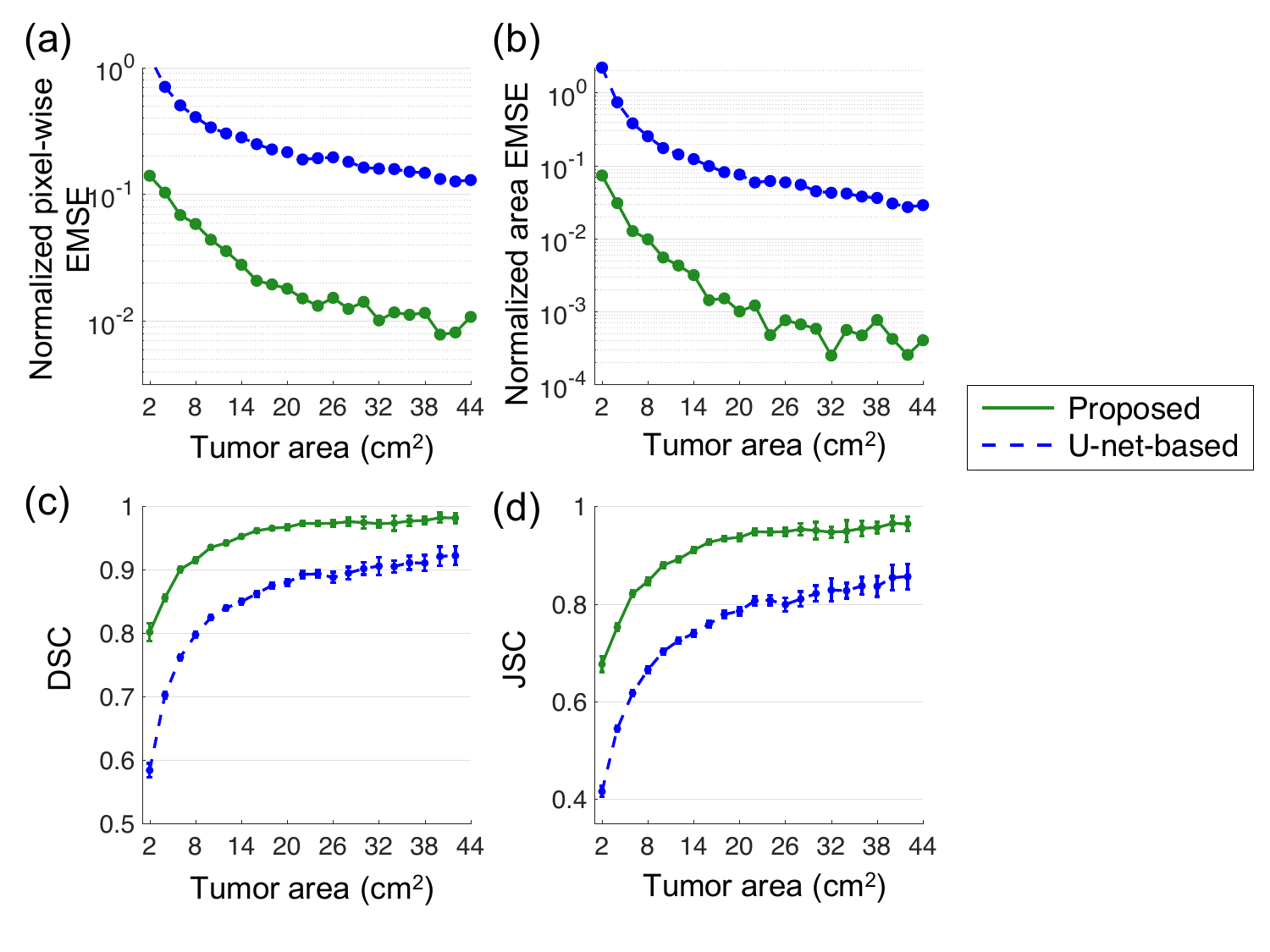}
    \caption{Evaluation result using clinically realistic simulation studies: effects of varying the tumor size on the task of (a) estimating the tumor-fraction areas, (b) estimating the whole tumor areas, and (c-d) segmenting the tumor. Plots (a-b) are displayed in log scale on y-axis for better visualization.}
    \label{fig:simul_tumor_area_exp}
\end{figure}

Fig. \ref{fig:simul_pve_exp} shows that the method yielded percent area overlap close to $100\%$ for all considered tumor sizes, including small tumors with axial cross-section less than $2~\mathrm{cm}^2$. For these smaller tumors, the diameter was approximately less than $3$ times the FWHM of the system resolution. This was unlike the PVEs-affected tumor areas, which, as expected, were significantly overestimated for smaller tumors. In addition, the proposed method yielded high DSC and JSC for these small tumors, indicating accurate segmentation performance. Further, the proposed method significantly outperformed the U-net-based method. Overall, these results demonstrate the relative insensitivity of the proposed method to PVEs when segmenting relatively small tumors. 
Further, Fig. \ref{fig:simul_tumor_area_exp} shows that the proposed method consistently yielded lower pixel-wise EMSE and lower area EMSE normalized by the true tumor areas, compared to the U-net-based method. The proposed method also yielded higher DSC and JSC for all tumor sizes.

\subsubsection{Evaluating accuracy of the proposed method for different clinical-scanner configurations}
\label{Sec: Evaluating accuracy of the proposed method using different clinical settings}
\begin{figure}[h]
    \centering
    \includegraphics[width=\textwidth]{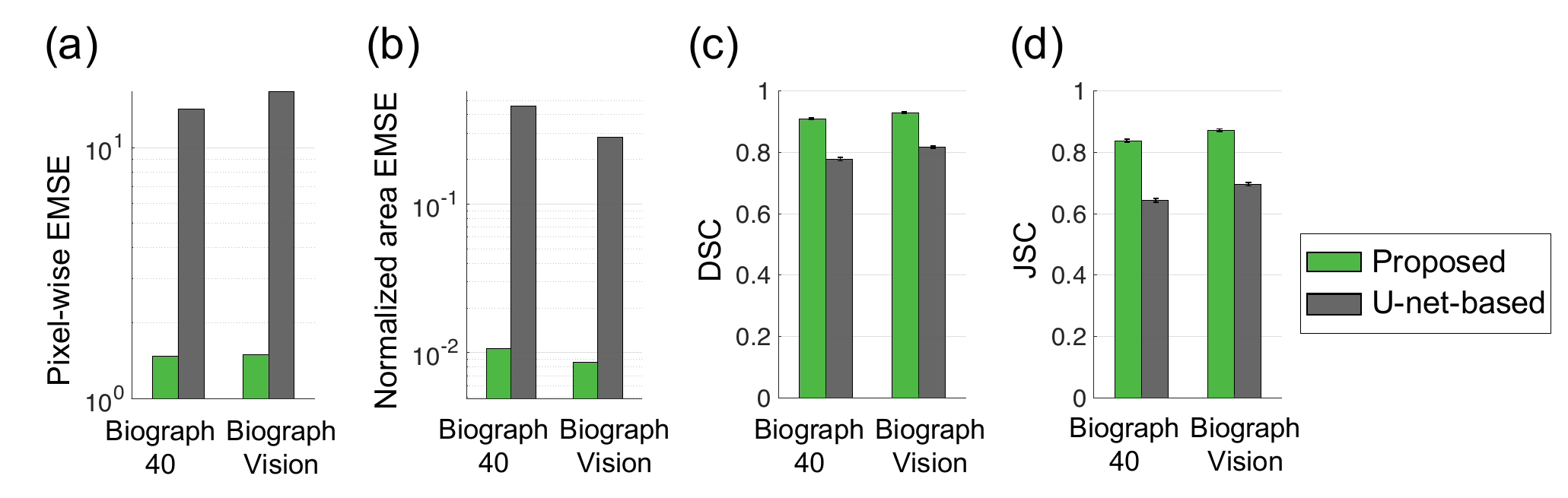}
    \caption{Evaluation result using clinically realistic simulation studies: evaluation of the segmentation performance for different clinical-scanner configurations on the basis of (a) pixel-wise EMSE, (b) normalized area EMSE, (c) Dice similarity coefficient, and (d) Jaccard similarity coefficient.}
    \label{fig:simul_clinical_scanner_exp}
\end{figure}

Fig. \ref{fig:simul_clinical_scanner_exp} shows the comparison of the segmentation accuracy between the proposed and the U-net-based method for two different clinical-scanner configurations, as described in Sec.~\ref{clinical configuration}. The proposed method significantly outperformed the U-net-based method for both clinical settings, on the basis of pixel-wise EMSE, normalized area EMSE, DSC, and JSC.

\subsection{Evaluation of the proposed method using clinical multi-center PET images}

Quantitatively, the proposed method yielded reliable segmentation with DSC of 0.82 ($95\%$ CI: $0.78,~0.86$). For 16 out of 17 test patients ($94.2\%$), both the proposed and U-net-based method yielded correct tumor localization in all $2$-D slices. When considering the patient cases with correct tumor localization, as shown in Fig.~\ref{fig:patient_metric_plot} (with details provided in Table~\ref{tab:patient_metric_table} in \ref{appendix a}), the proposed method significantly outperformed (p $ < 0.01$) all other considered methods, yielding the lowest pixel-wise EMSE, the lowest normalized area EMSE of 0.14, the highest DSC of $0.87$ ($95\%$ CI: $0.85,~0.89$), and the highest JSC of $0.74$ ($95\%$ CI: $0.70,~0.78$). In addition, the proposed method accurately segmented relatively small tumors and yielded high DSC of $0.77$ for the smallest segmented tumor axial cross-section of $1.30~\mathrm{cm}^2$ in area.

Qualitatively, we observe in Fig.~\ref{fig:patient_contour} that the proposed method yielded an accurate match to the true isocontours defined at different considered TFA levels, following the strategy in Sec.~\ref{sec:Evaluation on segmentation performance} with illustratration in Fig.~\ref{fig:isocontour_diagram}. 
Further, Fig.~\ref{fig:more_boundary_delineation} shows that the method accurately segmented tumors with small sizes (a, e), tumors with convex shape (b, f), tumors surrounded by regions with high uptake (c, d, g, h), and tumors with substantial intra-tumor heterogeneity (b, d, f, h).

\begin{figure}[h]
    \centering
    \includegraphics[width=0.95\textwidth]{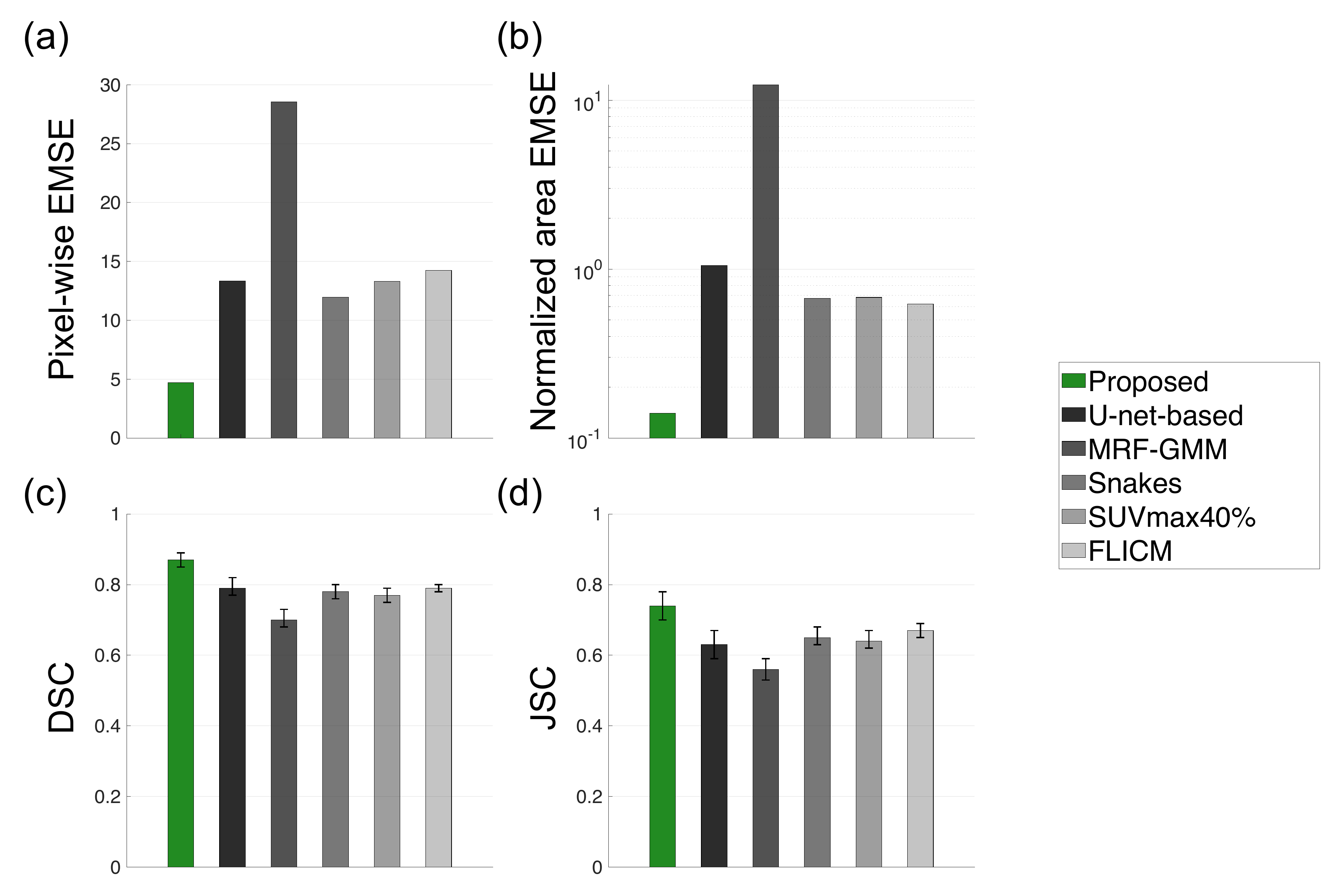}
    \caption{Evaluation result using clinical multi-center PET images: (a) the pixel-wise EMSE between the true and estimated tumor-fraction areas; (b) the normalized area EMSE between the measured and true tumor areas (plot displayed in log scale on y-axis for better visualization); the (c) Dice similarity coefficient and (d) Jaccard similarity coefficient between the true and estimated segmentations.}
    \label{fig:patient_metric_plot}
\end{figure}

\begin{figure}[h]
    \centering
    \includegraphics[width=\textwidth]{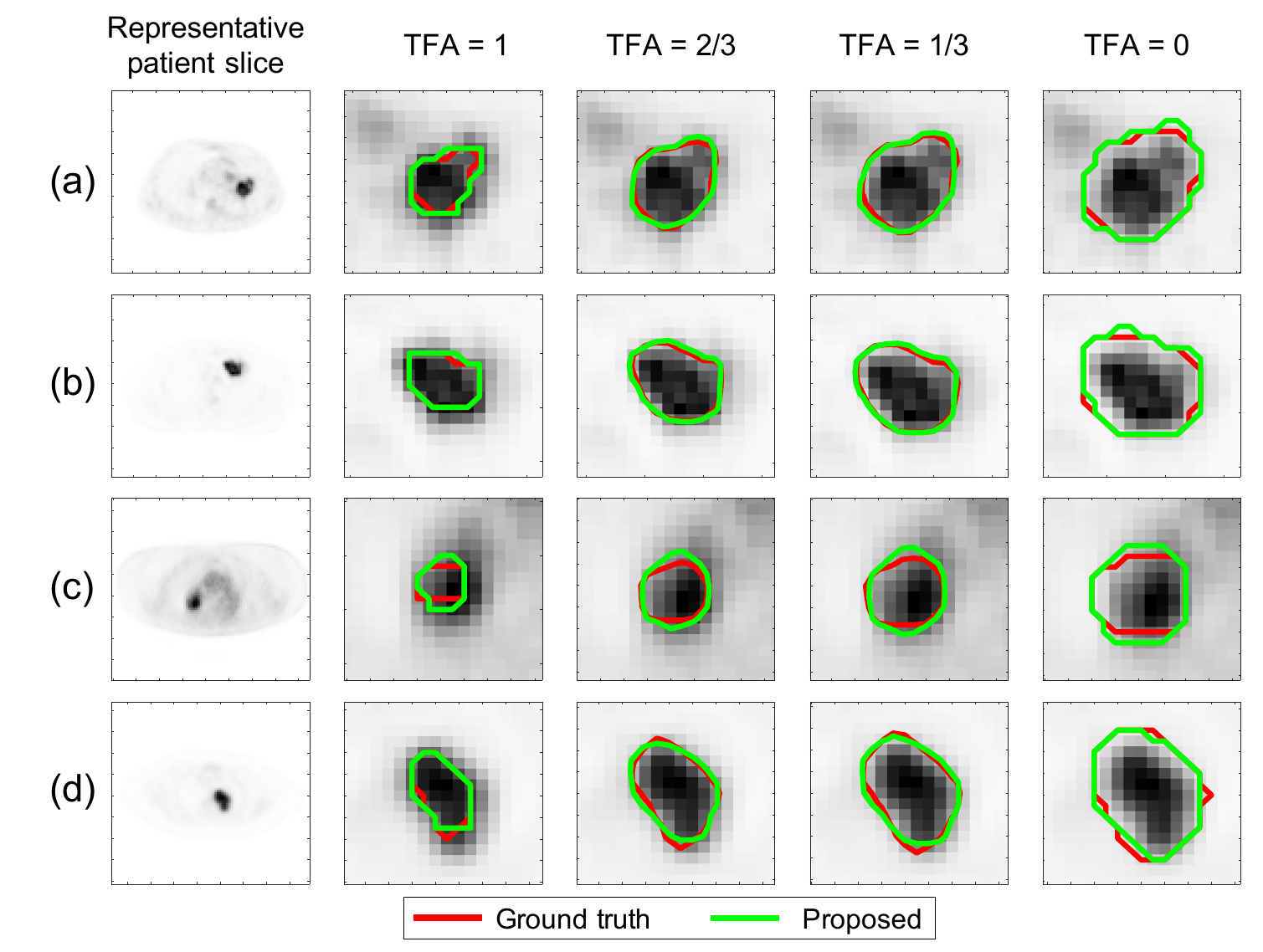}
    \caption{Evaluation result using clinical multi-center PET images: comparison between the estimated isocontours using the proposed method (green) and the ground-truth isocontours (red), defined as set of points at four TFA values (0, 1/3, 2/3, 1).}
    \label{fig:patient_contour}
\end{figure}

\section{Discussion}
\label{Sec: Discussion}

In this manuscript, we proposed a Bayesian approach to tissue-fraction estimation for segmentation in oncological PET. Conventional segmentation methods are typically classification-based, i.e.~classifying each voxel in the image as belonging to a certain tissue class. Thus, these methods are inherently limited in modeling TFEs. While probabilistic techniques can provide estimates of probabilities that each image voxel belongs to a tissue class, these probabilistic estimates are unrelated to TFEs. We address this inherent limitation by framing the segmentation task as an estimation problem, where the fractional volume that the tumor occupies in each voxel is estimated. Through this strategy, we are able to explicitly model the TFEs while performing segmentation. 

\begin{figure}[h!]
    \centering
    \includegraphics[width=\textwidth]{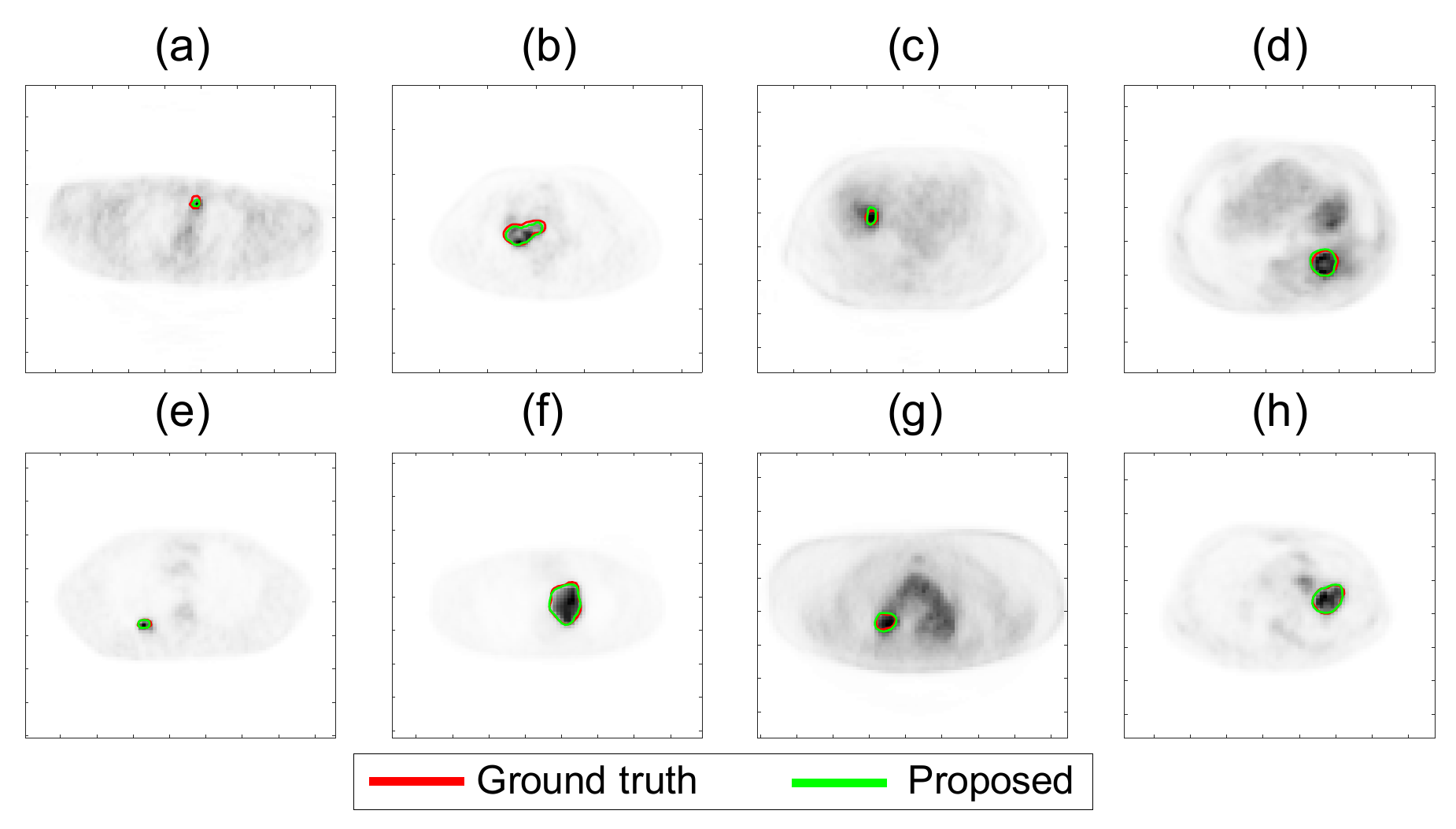}
    \caption{Evaluation result using clinical multi-center PET images: qualitative assessment of the performance of the proposed method in estimating the TFA maps for small tumors (a,e), for tumors with convex shape (b,f), for tumors surrounded by regions with high uptake (c-h), and for tumors with substantial intra-tumor heterogeneity (b,d,f,h). Isocontours were defined as set of points at TFA = $0.5$.}
    \label{fig:more_boundary_delineation}
\end{figure}

Quantitatively, the proposed method yielded accurate performance on estimation of the ground-truth TFA maps and on segmentation tasks, and significantly outperformed the considered segmentation methods, yielding the lowest pixel-wise EMSE and normalized area EMSE, and the highest DSC and JSC, as evaluated using both clinically realistic simulation studies (Fig.~\ref{fig:simul_metrics}) and clinical images from multi-center trial data (Fig.~\ref{fig:patient_metric_plot}). With clinical images, the method yielded a DSC of $0.82$ (95\% CI: $0.78$,~$0.86$). Qualitatively, the method yielded isocontours of close match to the ground-truth isocontours defined at different considered TFA values, as we observe from the results in Fig.~\ref{fig:simul_contour} and Fig.~\ref{fig:patient_contour}. Additionally, as shown in Fig.~\ref{fig:isocontour_diagram} for a representative tumor with substantial intra-tumor heterogeneity, the proposed method correctly estimates the TFA value as unity for pixels that are within the tumor boundary but have relatively low intensity. This observation was consistent across different heterogeneous tumors, showing the reliable performance of the proposed method even with heterogeneous tumors. We believe that the method is reliable in this scenario because the method estimates the TFA by computing the conditional expectation of the TFA in that pixel given the entire reconstructed PET image, and not just the intensity of that pixel (Eq.~\eqref{eq: vhatm*}). All these results demonstrate the ability of the method to accurately estimate the TFA within each image pixel and yield accurate tumor segmentations.

The isocontours defined based on certain choices of TFA values were shown only to visually illustrate the performance of the proposed method on the task of estimating the TFA map. The proposed method yields the estimated TFA map as the final output. This allows the method to provide the end user, such as a physician or a radiation oncologist, the ability to visualize the TFAs within each PET-image pixel, which they can use to make a decision based on their clinical use-case scenario.

Further, the proposed method demonstrated the ability to accurately segment relatively small tumors. In realistic simulation-based evaluation studies, the method yielded a high DSC of $0.84$ for the smallest segmented tumor, with an axial cross-section of $0.88~\mathrm{cm}^2$ and a diameter approximately twice the FWHM of the system resolution. With clinical images, for the smallest tumor axial cross-section of $1.30~\mathrm{cm}^2$, the method yielded a DSC of $0.77$. This accuracy in segmenting small tumors is especially important for clinical tasks such as radiotherapy planning, where an accurate segmentation for small tumors is crucial to protect normal organs from radiations.

While the U-net-based method had demonstrated the ability to account for PVEs arising due to the low system resolution \citep{leung2020physics}, the proposed method significantly outperformed this method, emphasizing the significance of modeling the TFEs in PET segmentation. This need to model TFEs was also demonstrated in the results of evaluation using clinically realistic simulation studies, where the performance of the method was assessed for different clinical-scanner configurations (Sec. \ref{Sec: Evaluating accuracy of the proposed method using different clinical settings}). For example, for the higher-resolution Biograph Vision scanner, the TFEs may be more dominant compared to system-resolution-related blur. We observed in Fig.~\ref{fig:simul_clinical_scanner_exp} that the proposed method was more accurate compared to the U-net-based method for this scanner. Further, for both clinical-scanner configurations, the proposed method yielded similar performance in estimating the TFAs and segmenting the tumor, indicating that the method was relatively insensitive to the changes in voxel size.  

Our evaluation of the proposed method with clinical images of patients with stage IIB/III NSCLC shows that the method, when trained with $61$ patients, yielded a reliable segmentation performance with DSC of $0.82$. When considering patient cases where the tumor was localized correctly by the method ($94.2\%$), the DSC further improved to $0.87$. These results demonstrate the accuracy of the method in clinical settings and motivate further clinical evaluation of the method with even larger datasets and with delineations defined by multiple readers. Further, the method is general, and the results motivate the evaluation of the method for segmenting tumors other than the primary tumors, including infiltrating tumors, and segmenting tumors at other stages of the disease, including metastasis. In all these cases, the method would require the corresponding definition of the ground-truth TFAs, or a surrogate for the ground truth, such as those from manual delineations performed by trained readers.

The results obtained with the proposed method also motivate further evaluation of this method for PET-based clinical applications that require tumor delineation such as PET-based radiotherapy planning \citep{el2009exploring,zaidi2009molecular}. Further, the results motivate evaluation of this method for the applications of computing PET-based volumetric markers of metabolic tumor volume (MTV) and total lesion glycolysis \citep{ohri2015pretreatment,chen2012prognostic}, and radiomic features \citep{cook2018challenges,zhang2017radiomics,mena201718f}, each of which are being evaluated as prognostic and predictive markers of therapy response. Such evaluation can be performed using task-specific evaluation frameworks \citep{kupinski2006comparing,jha2012task,jha2017,barrett2010therapy}. In this context, our initial results in both clinically realistic simulation (Fig.~\ref{fig:simul_metrics}(b)) and patient studies (Fig.~\ref{fig:patient_metric_plot}(b)) on estimating the tumor area indicate the promise of the proposed method on the task of quantifying MTV more accurately than conventional methods. 

Our study has some limitations. First, while the theory of the proposed method was developed in the context of $3$-D imaging, our evaluation studies were conducted on a per-slice basis. This helped to increase the size of training data and was computationally less expensive \citep{leung2020physics}. However, implementing the method to $3$-D segmentation is relatively straightforward and would require only slight modifications to our network architecture, such as the ability to be input $3$-D images and output $3$-D tumor-fraction volume maps. Thus, the $2$-D convolutional layers in the network would be replaced by $3$-D convolutional layers. The overall network design would remain similar. In fact, in the ongoing study on using an extended version of this method for segmenting $3$-D single-photon emission computed tomography (SPECT) images, we have seen that a similar design was sufficient to perform 3-D segmentation {\citep{HaeSol2020SPECT,liu2021fully}}. The results shown here and in the SPECT study suggest that the proposed method will yield reliable performance for $3$-D tumor segmentation in PET, and this is an area of further research. Additionally, in this study, the proposed method was used to segment the image into only two regions. However, the method is general, and in the ongoing study of 3-D SPECT segmentation, we are applying this method to segment the images into seven different regions. Another limitation is that our evaluation studies currently consider cases where only the primary tumor is present in an image.  However, again, the method could be generalized to potentially segment multiple tumors present in the same image slice. Confirming this though would require additional evaluation studies. Further, respiratory motion of the lung, which may also cause blurring of the tumor mask, was not considered in the proposed method. Extending the method to account for lung motion is also an important research area. Finally, the method does not incorporate tumor information from CT images while segmenting PET images. Incorporating information from CT images can provide a prior distribution of the tumor-fraction areas for the estimation task. Thus, investigating the incorporation of CT images into the proposed method is another important research direction.

We evaluated our method in the context of segmenting oncological PET images of patients with lung cancer and demonstrated accurate tumor segmentation performance. The method is general and thus, these results motivate the evaluation of the method for other cancer types. However, segmenting tumors in the lung region could be easier due to the scarce FDG uptake in the lung. In other cancer types, tumor-to-background intensity ratios may be lower, which may make the segmentation task challenging. For example, renal tumors often have similar FDG uptake as the normal renal cortex. Further, there may be situations where the FDG uptake in tumor is lower than the background, such as photon-deficient tumors on the liver. Thus, before application to other cancers, corresponding validation studies would be needed. Additionally, the method can be extended to segment PET images for other applications, such as those in cardiology and neurology. Further, the method can be extended to segment images from other imaging modalities that have low resolution, such as SPECT and optical imaging, with ongoing efforts in SPECT \citep{HaeSol2020SPECT,liu2021fully}.

\section{Conclusion}

In this manuscript, we proposed a Bayesian approach to tissue-fraction estimation for oncological PET segmentation. We theoretically demonstrated that the proposed method yields a posterior-mean estimate of the tumor-fraction volume for each voxel in the PET image. Evaluation of the method using clinically realistic $2$-D simulation studies demonstrated the capability of the method to explicitly model TFEs by accurately estimating the tumor-fraction areas. The method significantly outperformed the considered commonly used PET segmentation methods, including a U-net-based method. In addition, the method was relatively insensitive to partial-volume effects and demonstrated accurate segmentation performance for different clinical-scanner configurations. Further, the proposed method demonstrated accurate performance in segmenting clinical images of patients with stage IIB/III NSCLC, obtained from the ACRIN 6668/RTOG 0235 multi-center clinical trial data. For this dataset, the method yielded DSC of $0.82$ ($95\%$ CI: $0.78, 0.86$). In conclusion, this study demonstrates the efficacy of the proposed method for tumor segmentation in PET. Pending necessary permissions, we will publish the source code for the proposed method for wider usage by the image-science community (source code currently available at \url{https://drive.google.com/drive/folders/1Kk0LvnSUccz6zkYoKJgX73Wd9pvUTJm_?usp=sharing}).

\section*{Acknowledgments}
Financial support for this work was provided by the National Institute of Biomedical Imaging and Bioengineering R01 Award (R01-EB031051) and Trailblazer R21 Award (R21-EB024647), and the NVIDIA GPU grant. We also thank the Washington University Center for High Performance Computing for providing computational resources for this project. The center is partially funded by NIH grants 1S10RR022984-01A1 and 1S10OD018091-01.

\clearpage
\bibliography{ref}

\clearpage
\appendix
\section{}
\label{appendix a}

The architecture of the encoder-decoder network designed for the proposed method is provided in Table \ref{tab:network architecture}.

Details of the simulated PET systems used in the evaluation of proposed method for different clinical-scanner configurations are given in Table \ref{tab:TFE_PET_specs}.

Patient demographics with clinical characteristics and reconstruction parameters of clinical scanners in the ACRIN 6668/RTOG 0235 multi-center clinical trial are provided in Table \ref{tab:patient_demographics} and Table \ref{tab:scanner_info}, respectively.

Evaluation results of the proposed method using clinically realistic simulation studies and clinical images from multi-center clinical trial are given in Table~\ref{tab:simulation_metric_table} and Table~\ref{tab:patient_metric_table}, respectively.

\begin{table*}[h!]
\footnotesize
\captionsetup{justification=centering}
\caption{Architecture of the encoder-decoder network.}
\label{tab:network architecture}
\begin{tabularx}{\textwidth}{
    | >{\hsize=0.25\hsize\centering\arraybackslash}X
    | >{\hsize=0.50\hsize\centering\arraybackslash}X
    | >{\hsize=0.25\hsize\centering\arraybackslash}X
    | >{\hsize=0.17\hsize\centering\arraybackslash}X
    | >{\hsize=0.17\hsize\centering\arraybackslash}X
    | >{\hsize=0.29\hsize\centering\arraybackslash}X
    | >{\hsize=0.36\hsize\centering\arraybackslash}X|}
\hline
         & Layer Type       & Filter Size   & \# of Filters & Stride  & Input Size     & Output Size    \\ \hline
Layer 1  & Conv.            & 3$\times$3   & 32            & 1$\times$1      & 168$\times$168$\times$1  & 168$\times$168$\times$32 \\ \hline
Layer 2  & Conv.            & 3$\times$3   & 32            & 2$\times$2       & 168$\times$168$\times$32 & 84$\times$84$\times$32    \\ \hline
Layer 3  & Conv.            & 3$\times$3   & 64            & 1$\times$1       & 84$\times$84$\times$32    & 84$\times$84$\times$64    \\ \hline
Layer 4  & Conv.            & 3$\times$3   & 64            & 2$\times$2       & 84$\times$84$\times$64    & 42$\times$42$\times$64    \\ \hline
Layer 5  & Conv.            & 3$\times$3   & 128           & 1$\times$1       & 42$\times$42$\times$64    & 42$\times$42$\times$128   \\ \hline
Layer 6  & Conv.            & 3$\times$3   & 128           & 2$\times$2       & 42$\times$42$\times$128   & 21$\times$21$\times$128   \\ \hline
Layer 7  & Conv.            & 3$\times$3   & 256           & 1$\times$1       & 21$\times$21$\times$128   & 21$\times$21$\times$256   \\ \hline
Layer 8  & Conv.            & 3$\times$3   & 256           & 1$\times$1       & 21$\times$21$\times$256   & 21$\times$21$\times$256   \\ \hline
Layer 9  & Transposed Conv. & 3$\times$3   & 128           & 2$\times$2       & 21$\times$21$\times$256   & 42$\times$42$\times$128   \\ \hline
Layer 9 & Skip Connection (Add Layer 5)     & - & -             & -           & 42$\times$42$\times$128   & 42$\times$42$\times$128   \\ \hline
Layer 10 & Conv.            & 3$\times$3   & 128           & 1$\times$1       & 42$\times$42$\times$128   & 42$\times$42$\times$128   \\ \hline
Layer 11 & Transposed Conv. & 3$\times$3   & 64            & 2$\times$2       & 42$\times$42$\times$128   & 84$\times$84$\times$64    \\ \hline
Layer 11 & Skip Connection (Add Layer 3)      & - & -             & -           & 84$\times$84$\times$64    & 84$\times$84$\times$64    \\ \hline
Layer 12 & Conv.            & 3$\times$3   & 64            & 1$\times$1       & 84$\times$84$\times$64    & 84$\times$84$\times$64    \\ \hline
Layer 13 & Transposed Conv. & 3$\times$3   & 32            & 2$\times$2       & 84$\times$84$\times$64    & 168$\times$168$\times$32 \\ \hline
Layer 13 & Skip Connection (Add Layer 1)    & -  & -             & -           & 168$\times$168$\times$32 & 168$\times$168$\times$32 \\ \hline
Layer 14 & Conv.            & 3$\times$3   & 32            & 1$\times$1       & 168$\times$168$\times$32 & 168$\times$168$\times$32 \\ \hline
Layer 15 & Conv.            & 3$\times$3   & 2             & 1$\times$1       & 168$\times$168$\times$32 & 168$\times$168$\times$2  \\ \hline
Output & Softmax            & -   & -             & -       & 168$\times$168$\times$2 & 168$\times$168$\times$2  \\ \hline
\end{tabularx}
\end{table*}

\begin{table}[h]
    \centering
    \footnotesize
    \captionsetup{justification=centering}
    \caption{Technical acquisition and reconstruction parameters of the PET systems (FOV: field of view).}\label{tab:TFE_PET_specs}
    \begin{tabularx}{\textwidth}{
    | >{\hsize=0.8\hsize\centering\arraybackslash}X
    | >{\hsize=0.6\hsize\centering\arraybackslash}X
    | >{\hsize=0.6\hsize\centering\arraybackslash}X|}
    \hline
    Parameters & Biograph 40 & Biograph Vision \\
    \hline
    Transaxial FOV ($\mathrm{mm}$) & 550 & 700 \\ 
    \hline
    Axial FOV ($\mathrm{mm}$) & 216 & 260 \\
    \hline
    Reconstruction method & OSEM & OSEM \\
    \hline
    Subsets & 21 & 21 \\ 
    \hline
    Iterations & 2 & 2 \\ 
    \hline
    Crystal pitch ($\mathrm{mm}$) & 4.00 & 3.30 \\ 
    \hline
    FWHM ($\mathrm{mm}$) @ 1 $\mathrm{cm}$ & 5.90  & 3.70\\ 
    \hline
    Voxel size ($\mathrm{mm}^3$) & 4.30 $\times 4.30 \times 4.25$ & 3.65 $\times 3.65 \times 3.27$ \\
    \hline
    \end{tabularx}
\end{table}

\begin{table}[h]
    \centering
    \footnotesize
    \captionsetup{justification=centering}
    \caption{Patient demographics with clinical characteristics.}
    \label{tab:patient_demographics}
    \begin{tabularx}{\textwidth}{
    | >{\hsize=0.8\hsize\centering\arraybackslash}X
    | >{\hsize=0.6\hsize\centering\arraybackslash}X
    | >{\hsize=0.6\hsize\centering\arraybackslash}X|}
    \hline
    Demographics / clinical characteristics & Value & Percent \\
    \Xhline{3\arrayrulewidth}
    Age: median (range) & 67.5 (37 - 82) & - \\ 
    \Xhline{3\arrayrulewidth}
    Sex & Male & 63\% (49/78) \\
    \hline
     & Female & 37\% (29/78) \\
    \Xhline{3\arrayrulewidth}
    Race & White & 90\% (70/78) \\ 
    \hline
     & African American & 5\% (4/78) \\ 
    \hline
    & Asian & 2.5\% (2/78) \\ 
    \hline
    & Other/unknown & 2.5\% (2/78) \\ 
    \Xhline{3\arrayrulewidth}
    Performance status & Fully active & 41\% (32/78) \\ 
    \hline
     & Ambulatory & 59\% (46/78) \\ 
    \Xhline{3\arrayrulewidth}
    Clinical stage & IIB & 5\% (4/78) \\ 
    \hline
     & IIIA & 55\% (43/78) \\ 
    \hline
     & IIIB & 40\% (30/78) \\ 
    \Xhline{3\arrayrulewidth}
    Chemotherapy regimen & Carboplatin/paclitaxel & 60\% (47/78) \\
    \hline
     & Cisplatin/etoposide & 27\% (21/78) \\
    \hline 
     & Other & 12\% (9/78) \\
    \hline
     & Not available & 1\% (1/78) \\
    \Xhline{3\arrayrulewidth} 
    Radiation dose & $<$ 50 Gy & 1\% (1/78) \\
    \hline
     & 50-60 Gy & 8\% (6/78) \\
    \hline
     & 60-70 Gy & 58\% (45/78) \\
    \hline
     & $\geq$70 Gy & 27\% (21/78) \\
    \hline
     & Not available & 6\% (5/78) \\
    \Xhline{3\arrayrulewidth}
    \end{tabularx}
\end{table}

\begin{table*}[h!]
    \footnotesize
    \captionsetup{justification=centering}
    \caption{Reconstruction parameters of PET/CT systems used in ACRIN 6668/RTOG 0235 multi-center clinical trial. (DLYD: delayed event subtraction; SING: singles-based correction; N/A: not available)}
    \label{tab:scanner_info}
    \begin{tabularx}{\textwidth}{
    | >{\hsize=0.35\hsize\centering\arraybackslash}X
    | >{\hsize=0.33\hsize\centering\arraybackslash}X
    | >{\hsize=0.33\hsize\centering\arraybackslash}X
    | >{\hsize=0.33\hsize\centering\arraybackslash}X
    | >{\hsize=0.33\hsize\centering\arraybackslash}X
    | >{\hsize=0.33\hsize\centering\arraybackslash}X|}
    \hline
    Parameter &  GE Discovery ST & GE Discovery STE & GE Discovery RX & CPS 1023 & CPS 1024 \\
    \hline
    Reconstruction method
    & OSEM
    & OSEM
    & OSEM
    & OSEM
    & OSEM
    \\ \hline
    Subsets
    & N/A
    & N/A
    & N/A
    & 8
    & 8
    \\ \hline
    Iterations
    & N/A
    & N/A
    & N/A
    & 2
    & 2
    \\ \hline
    Attenuation correction
    & CT
    & CT
    & CT
    & CT
    & CT
    \\ \hline
    Scatter correction
    & Convolution subtraction
    & Convolution subtraction
    & Convolution subtraction
    & Model-based
    & Model-based
    \\ \hline
    Randoms correction
    & DLYD/SING
    & SING
    & SING
    & DLYD
    & DLYD
    \\ \hline
    Pixel spacing (mm) 
    & 4.69$\times$4.69 5.47$\times$5.47 
    & 5.47$\times$5.47
    & 5.47$\times$5.47
    & 5.31$\times$5.31
    & 5.31$\times$5.31
    \\ \hline
    Slice thickness (mm) 
    & 3.27 
    & 3.27
    & 3.27 
    & 2.50
    & 3.38
    \\ \hline
    \end{tabularx}
\end{table*}

\begin{table*}[h!]
    \footnotesize
    \captionsetup{justification=centering}
    \caption{Evaluation result using clinically realistic simulation studies: performance comparison between the proposed method and other considered segmentation methods.}\label{tab:simulation_metric_table}
    \begin{tabularx}{\textwidth}{
    | >{\hsize=0.35\hsize\centering\arraybackslash}X
    | >{\hsize=0.25\hsize\centering\arraybackslash}X
    | >{\hsize=0.30\hsize\centering\arraybackslash}X
    | >{\hsize=0.30\hsize\centering\arraybackslash}X
    | >{\hsize=0.25\hsize\centering\arraybackslash}X
    | >{\hsize=0.30\hsize\centering\arraybackslash}X
    | >{\hsize=0.25\hsize\centering\arraybackslash}X|}
    \hline
    Metrics &  Proposed & U-net-based & MRF-GMM & Snakes & $\mathrm{40\%~{SUV_{max}}}$ & FLICM \\
    \hline
    Pixel-wise EMSE & 2.04 & 17.00 & 25.64 & 43.63 & 27.13 & 30.26 \\
    \hline
    Normalized area EMSE & 0.02 & 0.49 & 4.07 & 8.55 & 1.34 & 1.78 \\
    \hline
    DSC & 0.90 (0.90,0.91) & 0.77 (0.77,0.78) & 0.73 (0.72,0.74) & 0.65 (0.64,0.66) & 0.64 (0.63,0.65) & 0.61 (0.60,0.62) \\
    \hline
    JSC & 0.83 (0.83,0.84) & 0.64 (0.64,0.65) & 0.60 (0.59,0.60) & 0.51 (0.50,0.52) & 0.50 (0.49,0.51) & 0.48 (0.47,0.49) \\
    \hline
    \end{tabularx}
\end{table*}

\begin{table*}[h!]
    \footnotesize
    \captionsetup{justification=centering}
    \caption{Evaluation result using clinical multi-center PET images: performance comparison between the proposed method and other considered segmentation methods on the basis of quantitative figures of merit. Results here are reported for patient cases with correct tumor localization (94.2\%).}
    \label{tab:patient_metric_table}
    \begin{tabularx}{\textwidth}{
    | >{\hsize=0.35\hsize\centering\arraybackslash}X
    | >{\hsize=0.25\hsize\centering\arraybackslash}X
    | >{\hsize=0.30\hsize\centering\arraybackslash}X
    | >{\hsize=0.30\hsize\centering\arraybackslash}X
    | >{\hsize=0.25\hsize\centering\arraybackslash}X
    | >{\hsize=0.30\hsize\centering\arraybackslash}X
    | >{\hsize=0.25\hsize\centering\arraybackslash}X|}
    \hline
    Metrics &  Proposed & U-net-based & MRF-GMM & Snakes & $\mathrm{40\%~{SUV_{max}}}$ & FLICM \\
    \hline
    Pixel-wise EMSE & 4.70 & 13.33 & 28.55 & 13.30 & 14.23 & 11.97 \\
    \hline
    Normalized area EMSE & 0.14 & 1.05 & 12.33 & 0.68 & 0.62 & 0.67 \\
    \hline
    DSC & 0.87 (0.85,0.89) & 0.79 (0.77,0.82) & 0.70 (0.68,0.73) & 0.78 (0.76,0.80) & 0.77 (0.75,0.79) & 0.79 (0.78,0.80) \\
    \hline
    JSC & 0.74 (0.70,0.78) & 0.63 (0.59,0.67) & 0.56 (0.53,0.59) & 0.65 (0.63,0.68) & 0.64 (0.62,0.67) & 0.67 (0.65,0.69) \\
    \hline
    \end{tabularx}
\end{table*}

\clearpage

\section{}
\label{appendix b}
\setcounter{equation}{0}
\renewcommand{\theequation}{B\arabic{equation}}

In this appendix, we provide the proof of showing that the optimal estimator minimizing the cost function in Eq.~\eqref{eq: Cost function final form} is unbiased in a Bayesian sense. 
To show this, we take the average of the estimate $\hat{\mathbf{v}}^*$ over the joint distribution of noise realizations $\hat{\mathbf{f}}$ and true values $\mathbf{v}$:
\begin{equation}
\begin{split}
    \overline{\overline{\hat{\mathbf{v}}^*}} 
    &= \int d^M\mathbf{v} \int d^M \mathbf{\hat{f}} \ \mathrm{pr}(\mathbf{\hat{f}},\mathbf{v}) \hat{\mathbf{v}}^* \\ 
    &= \int d^M\mathbf{v} \ \mathrm{pr}(\mathbf{v}) \int d^M \mathbf{\hat{f}} \  \mathrm{pr}(\mathbf{\hat{f}}|\mathbf{v}) \hat{\mathbf{v}}^* \\
    &= \int d^M\mathbf{v} \ \mathrm{pr}(\mathbf{v}) \int d^M \mathbf{\hat{f}} \  \mathrm{pr}(\mathbf{\hat{f}}|\mathbf{v}) \int d^M\mathbf{v'} \ \mathrm{pr}(\mathbf{v'}|\mathbf{\hat{f}}) \mathbf{v'},
\end{split}
\end{equation}
where in the second step we have expanded $\mathrm{pr}(\mathbf{\hat{f}},\mathbf{v})$ using the conditional probability, and in the third step we have inserted Eq. \eqref{eq: vhat*}. By using the Bayes' theorem and changing the order of integration, the above equation becomes
\begin{equation}
    \overline{\overline{\hat{\mathbf{v}}^*}} = \int d^M \mathbf{\hat{f}} \ \mathrm{pr}(\mathbf{\hat{f}}) \int d^M\mathbf{v'} \ \mathrm{pr}(\mathbf{v'}|\mathbf{\hat{f}}) \mathbf{v'} \int d^M \mathbf{v} \  \mathrm{pr}(\mathbf{v}|\mathbf{\hat{f}}).
\label{eq: ensemble average step 1}
\end{equation}
Since $\int d^M\mathbf{v} \ \mathrm{pr}(\mathbf{v}|\mathbf{\hat{f}}) = 1$, Eq. \eqref{eq: ensemble average step 1} becomes
\begin{equation}
    \overline{\overline{\hat{\mathbf{v}}^*}}  = \int d^M \mathbf{\hat{f}} \ \mathrm{pr}(\mathbf{\hat{f}}) \int d^M\mathbf{v'} \ \mathrm{pr}(\mathbf{v'}|\mathbf{\hat{f}}) \mathbf{v'}.
\end{equation}
Further, we can simplify the above equation using the law of total expectation and get
\begin{equation}
    \overline{\overline{\hat{\mathbf{v}}^*}} = \int d^M\mathbf{v'} \ \mathrm{pr}(\mathbf{v'}) \mathbf{v'} = \overline{\mathbf{v}}.
    \label{eq: unbiased in Bayesian sense}
\end{equation}
Thus, the average value of the estimate is equal to the average true value, so that the estimator is unbiased in a Bayesian sense. 

\end{document}